\documentclass[longauth]{aa}

\usepackage{graphicx}
\usepackage{placeins}
\usepackage{txfonts}
\newcommand{\micron}{$\mu$m}              %solar luminosities    

\usepackage[colorlinks=true,linkcolor=bluea,allcolors=blue]{hyperref}

\begin{document}

%   \title{Constraining the torus properties of a type-2 quasar with silicate features in emission using JWST MIRI spectra}

% Chat-GPT suggestions for titles
%\title{The Galaxy Activity, Torus, and Outflow Survey (GATOS): N. Local predictions on the infrared emission of type-1 AGN from JWST}
\title{The Galaxy Activity, Torus, and Outflow Survey: N. Little Red Dots are not dust deficient: constraints from Seyfert 1 JWST spectra}

%Unveiling Hot Dust in a Type-2 Quasar with JWST: Evidence for Low Torus Covering Factors and Radiative Blowout

%Probing the Dusty Torus of a Type-2 Quasar at z ∼ 0.1 with JWST/MIRI and Bayesian Torus Models

 %  \subtitle{I. Overviewing the $\kappa$-mechanism}

   \author{C. Ramos Almeida\inst{1,2}, 
%          D. Esparza-Arredondo\inst{1,2},
%          O. Gonz\'alez-Mart\'in\inst{3}, 
R. Maiolino\inst{3,4,5}, 
I. García-Bernete\inst{6},
A. Alonso-Herrero\inst{6},
L. Zhang\inst{7},
M. Pereira-Santaella\inst{8},
A. Asensio Ramos\inst{1,2},
S. Hönig\inst{9},
N. A. Levenson\inst{10},
P. G. Pérez-González\inst{3},
O. Gonz\'alez-Mart\'in\inst{11}, 
D. Rosario\inst{12},
R. Davies\inst{13},
X. Ji\inst{3,4},
E. Lopez-Rodriguez\inst{14},
A. Audibert\inst{1,2},
F. Combes\inst{15},
E. Bellocchi\inst{16,17}, 
P. Boorman\inst{13},
A. Bunker\inst{18},
L. Colina\inst{6},
T. Diaz-Santos\inst{19},
F. Donnan\inst{20},
S. García-Burillo\inst{21},
B. García-Lorenzo\inst{1,2},
L. Hermosa Mu{\~n}oz\inst{22}, 
E. K. S. Hicks\inst{23},
M. Imanishi\inst{24,25}
A. Labiano\inst{3,26},
C. Ricci\inst{27,28,29},
D. Rigopoulou\inst{19},
\and
M. Stalevski\inst{30,31}
}

   \institute{Instituto de Astrof\' isica de Canarias, Calle V\' ia L\'actea, s/n, E-38205, La Laguna, Tenerife, Spain\\
              \email{cra@iac.es}
              \and Departamento de Astrof\' isica, Universidad de La Laguna, E-38206, La Laguna, Tenerife, Spain
\and Kavli Institute for Cosmology, University of Cambridge, Madingley Road, Cambridge, CB3 0HA, UK
\and Cavendish Laboratory - Astrophysics Group, University of Cambridge, 19 JJ Thomson Avenue, Cambridge, CB3 0HE, UK
\and Department of Physics and Astronomy, University College London, Gower Street, London WC1E 6BT, UK
\and Centro de Astrobiolog\' ia (CAB), CSIC-INTA, Camino Bajo del Castillo s/n, E-28692, Villanueva de la Cañada, Madrid, Spain
\and Department of Physics and Astronomy, The University of Texas at San Antonio (UTSA), 1 UTSA Circle, San Antonio, TX 78249-0600, USA
\and Instituto de Física Fundamental, CSIC, Calle Serrano 123, 28006 Madrid, Spain
%\and Kavli Institute for Particle Astrophysics \& Cosmology (KIPAC), Stanford University, Stanford, CA 94305, USA
%\and Department of Physics \& Astronomy, University of South Carolina, Columbia, SC 29208, USA 
\and School of Physics \& Astronomy, University of Southampton, Highfield, Southampton SO171BJ, UK
\and Space Telescope Science Institute, Baltimore, MD 21218, USA 
\and Instituto de Radioastronom\' ia and Astrof\' isica (IRyA-UNAM), 3-72 (Xangari), 8701, Morelia, Mexico
\and School of Mathematics, Statistics and Physics, Newcastle University, New- castle upon Tyne NE1 7RU, UK
\and Max Planck Institute for Extraterrestrial Physics (MPE), Giessenbachstr. 1, D-85748 Garching, Germany
\and Department of Physics \& Astronomy, University of South Carolina, Columbia, SC 29208, USA
\and LUX, Observatoire de Paris, Collège de France, PSL University, CNRS, Sorbonne University, Paris, 75014, France
\and Depto. de Física de la Tierra y Astrofísica, Facultad de CC. Físicas, Universidad Complutense de Madrid, E-28040 Madrid, Spain
\and Instituto de Física de Partículas y del Cosmos IPARCOS, Facultad CC. Físicas, Universidad Complutense de Madrid, E-28040 Madrid, Spain
\and Department of Physics, University of Oxford, Oxford OX1 3RH, UK
\and Institute of Astrophysics, Foundation for Research and Technology-Hellas (FORTH), Heraklion, GR-70013, Greece
\and Department of Astrophysics, University of California San Diego, 9500 Gilman Drive, San Diego, CA 92093, USA
\and Observatorio Astron\'omico Nacional (OAN-IGN)-Observatorio de Madrid, Alfonso XII, 3, 28014 Madrid, Spain 
\and Departamento de Física, Universidad de Oviedo, Campus de Llamaquique, C/ Calvo Sotelo s/n, 33007 Oviedo, Spain
\and Department of Physics \& Astronomy, University of Alaska Anchorage, Anchorage, AK 99508-4664, USA
\and National Astronomical Observatory of Japan, National Institutes of Natural Sciences (NINS), 2-21-1 Osawa, Mitaka, Tokyo 181-8588, Japan
\and Department of Astronomy, School of Science, Graduate University for Advanced Studies (SOKENDAI), Mitaka, Tokyo 181-8588, Japan
%\and Kavli Institute for Cosmology, University of Cambridge, Madingley Road, Cambridge, CB3 0HA, UK
%\and Cavendish Laboratory - Astrophysics Group, University of Cambridge, 19 JJ Thomson Avenue, Cambridge, CB3 0HE, UK
%\and Department of Physics and Astronomy, University College London, Gower Street, London WC1E 6BT, UK
%\and Department of Physics \& Astronomy, University of Sheffield, S3 7RH Sheffield, UK
%\and Instituto de Astrof\' isica de Andaluc\' ia – CSIC, Glorieta de la Astronomía s/n, 18008 Granada, Spain
\and Telespazio UK for the European Space Agency, ESAC, Camino Bajo del Castillo s/n, 28692 Villanueva de la Cañada, Spain
%\and Department of Physics \& Astronomy, University of Sheffield, S3 7RH Sheffield, UK
%
\and Department of Astronomy, University of Geneva, ch. d'Ecogia 16, 1290, Versoix, Switzerland
\and Instituto de Estudios Astrof\'{\i}sicos, Facultad de Ingenier\'{\i}a y Ciencias, Universidad Diego Portales, Avenida Ej\'ercito Libertador 441, Santiago, Chile
\and Kavli Institute for Astronomy and Astrophysics, Peking University, Beijing 100871, China
\and Astronomical Observatory, Volgina 7, 11060 Belgrade, Serbia
\and Department of Physics and Astronomy, Universiteit Gent, Proeftuinstraat 86 N3, B-9000 Ghent, Belgium
}       

   \date{Received Nov 19, 2026; accepted XXX XX, 2026}

% \abstract{}{}{}{}{} 
% 5 {} token are mandatory
 
  \abstract
{

Little Red Dots (LRDs) are compact, red sources discovered by the James Webb Space Telescope (JWST). Most show broad permitted emission lines characteristic of active galactic nuclei (AGN), but other properties differ from local AGN, such as X-ray and radio weakness and an apparent scarcity of dust. If LRDs at $z\gtrsim4$ are indeed dust deficient, the bolometric luminosities and black hole (BH) masses measured from spectral energy distribution (SED) fitting would be overestimated, and their red optical continuum would be intrinsic rather than reddened. Here we use rest-frame 3--27\,$\mu$m JWST spectra of a sample of nearby ($D<60$\,Mpc) Seyfert~1 (Sy1) nuclei to place constraints on the dust emission of LRDs. We normalised the individual Sy1 spectra by their broad H$\alpha$ and 5100\,\AA\ continuum luminosities and stacked them, obtaining $\lambda L_\lambda/L_{\rm H\alpha}\simeq30$--60 and $\lambda L_\lambda/\lambda L_{5100}\simeq0.8$--1.8 across 3--27\,$\mu$m. These scalings reproduce, within a factor of three, the rest-frame near-infrared emission ($\lambda\lesssim4$\,$\mu$m) of a NIRSpec+MIRI stack of 249 LRDs at $2.3<z<9.3$ and the near- and mid-infrared ($\lambda\lesssim22$\,$\mu$m) emission of a local LRD at $z=0.1$, with the 5100\,\AA\ continuum giving the more reliable prediction when modest extinction is considered ($A_V\lesssim1$\,mag). However, simultaneously reproducing the optical and infrared SEDs of LRDs requires a more reddened accretion disc ($A_V\simeq2$\,mag) and a torus with a lower reprocessing efficiency ($C_{\rm T}\simeq0.1-0.2$) than inferred for local Sy1 galaxies. This combination offers an alternative to BH-star scenarios that explain the optical continuum as an unreddened blackbody of $T\sim5000$\,K. Our model, based on local Sy1s, yields a more plausible infrared-to-optical energy balance, suggesting that LRDs are not dust deficient. From the torus model that reproduces the optical and near-infrared SED of the high-redshift LRD stack, we estimate a dust mass of $M_{\rm dust}\simeq10^4\,M_\odot$, consistent with the dust masses typical of AGN tori.

%Indeed, from the torus model that reproduces the optical and near-infrared SED of the high-redshift LRD stack, we estimate a dust mass of $M_{\rm dust}\simeq10^4\,M_\odot$. %This dust mass is two orders of magnitude below the current upper limits for LRDs, but higher than the dust masses of local Seyfert galaxies 

%Using these normalizations and a NIRSpec+MIRI stack of 249 LRDs at 2.3<z<9.3, we find that the near-infrared emission predicted from broad H$\alpha$ is a factor $\sim$1-3 higher than observed (without any extinction correction). This could imply a higher covering factor broad line region in LRDs compared to local AGN, but lower than previously claimed. In the case of the infrared emission predicted from the 5100 \AA~continuum, it matches the observed one considering a small extinction of $\rm A_V\lesssim$1 mag. We find the same results when we predict the infrared emission of a local LRD at z=0.1. 
%Using a  from which we measure the broad H$\alpha$ and 5100 \AA~fluxes, we find that both quantities successfully predict its near-infrared emission, considering low extinction values of $\rm A_V\sim$0-1 mag. 
%We conclude that the infrared emission of LRDs is not that different from those of local Seyfert 1 galaxies with bolometric luminosities between 42 and 44.5 erg~s$^{-1}$.
%{\color {magenta} To be finished.}
}

%the infrared emission expected from the broad H$\alpha$ and 5100 \AA~fluxes of high-redshift LRDs is similar to that of local Sy1 galaxies with bolometric luminosities of $\sim$42-44.5 erg~s$^{-1}$.} 

   
\titlerunning{LRDs are not dust deficient}
\authorrunning{C. Ramos Almeida et al.}

   \maketitle
%
%-------------------------------------------------------------------

\section{Introduction}
\label{intro}

The study of active galactic nuclei (AGN) has become increasingly compelling with the advent of the James Webb Space Telescope (JWST). This is particularly the case for AGN at redshift z$\gtrsim$4 \citep{Kocevski2023,Harikane2023,Adamo2025}, which are no longer restricted to bright quasars ($\rm L_{bol}\gtrsim~10^{46}~erg~s^{-1}$). JWST/NIRSpec spectroscopy has revealed a large population of broad-line AGN with $\rm L_{bol}\simeq~10^{44-45}~erg~s^{-1}$ and black hole (BH) masses of $\rm M_{BH}\simeq~10^{6-8}~M_{\sun}$ (e.g. \citealt{Maiolino24,Kocevski25,Taylor2025,Geris2026,Juodzbalis2026,Perez26}). These AGN luminosities and BH masses are now closer to the average values of local active galaxies (see Table \ref{tab1}), which now, thanks to JWST/MIRI, can be studied in the mid-infrared with unprecedented spectral resolution and coverage \citep{Zhang24a,GarciaBernete24,Gonzalez25,Ramos25,Hermosa2025,Donnan2026,Goold2026}. 

Interestingly, the population of AGN at z$\gtrsim$4 discovered by JWST show some properties that are substantially different from those of nearby AGN. Despite the broad Balmer lines detected in a significant fraction of them, indicating that they are type-1 AGN, they are extremely weak in the X-rays \citep{Ananna24,Yue2024,Maiolino25,Tortosa2026,Geris2026b}. They are also faint at radio frequencies \citep{Mazzolari2024,Mazzolari2025}, and they show weak or no significant ultraviolet (UV) to optical variability \citep{Kokubo2025,Tee2025}, although this aspect has been debated \citep{Furtak2025,Ji25a,Lambrides2026,Lin2026variab,Secunda2026}.

While many of the broad line AGN found by JWST have blue colors (so-called Little Blue Dots, \citealt{Hainline2025,Brazzini26,Geris2026b}), a fraction of them (which depends on luminosity and selection band) is characterized by red optical slope and blue UV slope, giving them a characteristic ``V-shaped'' continuum -- these have been named ``Little Red Dots'' \citep{Matthee24,Greene2024,Kocevski25}. In addition to the several peculiar properties discussed above, LRDs are often characterized by Balmer absorption lines \citep{Matthee24,Ignas24,Ignas2026b,DEugenio2025,Matthee2026} and also Balmer breaks. The latter were initially interpreted as tracing evolved stellar populations, resulting in extremely large stellar masses (possibly cosmology-breaking; \citealt{Inayoshi2024}) and extremely high stellar densities \citep{Baggen2024}. However, subsequent studies have revealed that such features are tracing dense gas along the line of sight \citep{Inayoshi2025,Ji25a,Naidu2025,deGraaf2025a}. The nature of such absorbing gas along the line of sight is matter of debate, with scenarios ranging from Broad Line Region (BLR) clouds to a low-dust-content torus \citep{Ignas24,Madau2026b,Zhong2026}, to a partially ionized cocoon \citep{Rusakov2026,Sneppen2026}, and to an embedding pseudo-stellar atmosphere \citep{Naidu2025,Liu2025}.

Another important question is the origin of the red optical shape of LRDs. Many authors suggest that the optical continuum redward of the Balmer break resembles a black body (BB) with temperature around 5000 K \citep{deGraaff2025,Lin25}, and this is interpreted as the signature of a pseudo-stellar photosphere \citep{Liu2025,Kido2025}. %However, the optical spectrum does not actually match a BB and needs to be modified with a wavelength dependent factor \citep{deGraaf2025a}. 
\citet{Sneppen2026a} ascribe the optical spectral shape to nebular emission, based on the detection of a Paschen jump in a few LRDs. Other authors interpret the optical shape in terms of dust reddening, possibly associated with the classical torus \citep[e.g.][]{Madau2026b}, but with lower dust-to-gas ratio than in typical AGN, owing the lower metallicity typical of LRDs ($\sim 0.1~Z_\odot$, consistent with the average metallicity normal galaxies at the same cosmic epoch).

The scenario of dust absorption was initially discarded because of early claims that LRDs show no or little evidence for dust emission \citep{Williams2024,Akins2025}. The non-detections with the Atacama Large Millimeter/submillimeter Array (ALMA), which provide upper limits on the dust mass down to $\sim10^6~M_\odot$ \citep{Casey25,Greene2026,Akins2025}, are often taken as evidence of lack of dust. However, one should take into account that the dust mass of the nuclear torus is typically very low: $\rm \sim6\times 10^3~M_{\odot}$ as measured from high-angular resolution ALMA observations of local Seyfert galaxies \citep{GarciaBurillo21}, and between $\rm \sim~10^2$ and $\rm 10^4~M_{\odot}$ from SED fitting of SDSS quasars at z$\leq$0.5 \citep{Hatziminaoglou2008}. For what concerns the host galaxy, the vast majority of distant, low metallicity galaxies are undetected with ALMA - therefore the ALMA non-detection of LRDs are not surprising, but simply consistent with the fact that LRDs are hosted in low-metallicity and low-mass galaxies. Additionally, sub-mm non-detections at high-redshift should be interpreted cautiously, because the increasing temperature of the Cosmic Microwave Background (CMB) reduces the contrast between dust and the CMB emission \citep{daCunha2013}.

Interestingly, more recent works have found evidence for hot dust emission (T$\sim$1000-1500 K), typical of AGN, in the rest-frame near-infrared spectrum of many LRDs, both individually and in stacks \citep{Ignas24,Delvecchio2025,Lin2026,Brazzini26,Perez26}. Obviously, the detection of this hot dust emission has been more difficult at high-z, because the thermal bands are shifted outside JWST's sensitive bands, which required the use of stacking, but in LRDs at z$\lesssim$3 hot dust is clearly detected \citep{Ignas24,Ji2026,Lin25,Lin2026,Brazzini26}. In the case of local LRDs (z$\lesssim$0.5), for which there are mid-infrared SEDs available \citep{Ji2026,Lin25}, dust at temperatures of T$\sim$100-300 K, and up to T$\sim$1000 K, have been reported. The question is whether the dust emission of LRDs is drastically different from that of normal AGN or not. Analysing the Red Rosetta Stone, an LRD at z=2.6, \cite{Brazzini26} find a deficiency of mid-infrared emission by a factor ranging from 3 to 6 relative to the continuum at 1 $\mu$m and using the local AGN templates from \citet{GarciaBernete19}. \citet{Madau2026b} show that the LRD stack by \cite{Delvecchio2025}, as well as the spectrum and photometry of The Cliff LRD \citep{deGraaf2025a}, can be modelled in the framework of the simple AGN unified model (though with a high accretion rate), with a dusty torus covering factor of C$_T\sim$0.2.

These various results indicate that
LRDs might not be as dust-deficient as initially claimed.
More generally, the understanding of dust emission in LRDs is far from settled, and this can have profound implications for the scenarios invoked to explain their nature.

In this paper, we explore more thoroughly the dust emission of LRDs. We start by re-assessing the nuclear mid-infrared emission of local type-1 AGN. While this was already explored in the past based on ground-based and low resolutions space facilities, JWST now offers detailed mid-infrared information on the dust and nebular emission of local AGN by isolating the nuclear component from circumnuclear emission associated with stars and star formation. Additionally, we derive the scaling between the infrared and optical emission in type-1 AGN. This provides a more careful comparison to infer the amount of dust emission expected in AGN at high-z, especially with luminosities not as extreme as those of powerful quasars. We use these local templates to revisit the dust emission in distant and local LRDs. As we will illustrate, this analysis indicates that LRDs are not dust deficient, although their torus properties differ from those of local AGN. In the following, we assume a cosmology with H$_0$=70 km~s$^{-1}$ Mpc$^{-1}$, $\Omega_m$=0.3, and $\Omega_{\Lambda}$=0.7.

%Specifically, their mir-IR is consistent with dust in the circu

% At least $\sim$20\% of the AGN at z$\gtrsim$4-7 detected with JWST show absorption features on the Balmer emission line profiles \citep{Maiolino24,Matthee24,Ignas24,Kocevski25}. This has been interpreted as absorption by extremely dense gas ($\rm n_H\gtrsim10^9~cm^{-3}$) along the line-of-sight \citep{Inayoshi2025,Naidu2025,Ji25a}, explaining the presence of broad lines and possibly the weak X-ray emission (although even higher column densities would be required to explain the latter because of the low metallicities). Balmer absorption in the local universe is rare ($\sim$0.05\% occurrence at z<0.35 according to the recent study of \citealt{Shangguan2026}), likely due to the action of AGN feedback. 

\begin{table*}[h!]
\centering
\footnotesize
  \caption{Properties of the Seyfert 1 galaxies.} 
\begin{tabular}{lcccccccccc}
\hline
\hline
ID   &  Seyfert & Dist & Scale & i & $\rm f_{\rm c}$ & $\rm log~N{\rm _H}$ & $\rm log~L_{\text{2-10 keV}}$ & $\rm log~L_{\rm {bol}}$ & $\rm log~ M_{\rm BH}$  & log\(\frac{L_{\rm bol}}{L_{\rm Edd}}\) \\
  & & (Mpc)  & (pc~arcsec$^{-1}$) & (deg) & & (cm$^{-2}$) & (erg s$^{-1}$) & (erg s$^{-1}$)  & (M$_{\odot}$)  &  \\
\hline
NGC1097  & 1   & 14.5 (a)        & 70  & 50 & 0.15 & 20.60 & 40.90 & 41.90 & 8.10 & -4.30 \\  
NGC3227  & 1.5 & 23.7 (b)        &115  & 56 & 0.40 & 20.95 & 42.10 & 43.40 & 6.58 & -1.28 \\
NGC4051  & 1.5 & 16.6 (c$^{\S}$) & 80  & 35 & 0.80 & 20.00 & 41.33 & 42.63 & 5.95 & -1.42 \\
NGC4395  & 1.5 & 4.30 (d)        & 21  & 40 & 0.74 & 21.04 & 40.50 & 41.80 & 4.43 & -0.73 \\
NGC6814  & 1.2 & 21.7 (e)        & 105 & 17 & 0.51 & 20.97 & 42.31 & 43.61 & 6.89 & -1.38 \\
NGC7213  & 1.5  & 22.0 (a)       & 107 & 20 & 0.12 & 20.00 & 41.95 & 43.25 & 6.41 & -1.26 \\
NGC7314  & 1.8/1.9$^{*}$&18.3 (a)&  89 & 64 & 0.84 & 21.60 & 42.33 & 43.63 & 6.70 & -1.17 \\
NGC7469  & 1 & 58.0 (f)          & 281 & 45 & 0.40 & 20.53 & 43.19 & 44.49 & 7.48 & -1.09 \\ 
\hline 
\end{tabular}
\tablefoot{Galaxy inclination (i) are from NED outer diameters and geometrical torus covering factor ($\rm f_c$) derived by \citet{Ramos11b} and \citet{GarciaBernete19} from infrared SED fitting using CLUMPY torus models \citep{Nenkova08a,Nenkova08b}. The column density and intrinsic 2-10 keV luminosities are from \citet{Ricci17b} and M$_{\rm BH}$ from the H$\alpha$ measurements reported in \citet{Mejia22}, assuming a virial factor of f = 1. The only exceptions are NGC\,1097, whose values come from \citet{Cisternas13} and \citet{Lewis06} respectively, and NGC\,7314, whose BH mass is from \citet{Schulz1994}. $L_{\rm bol}$ is calculated as 20$\times$L$_{\rm 2-10 keV}$ except for NGC\,1097, for which we used 10$\times$L$_{\rm 2-10 keV}$ \citep{Ho2009}, and L$_{\rm Edd}$=1.26$\times$10$^{38}\times$M$_{\rm BH}$. $^{*}$ Usually classified as Seyfert 1.8/1.9 but broad H$\alpha$ and H$\beta$ components reported by \citet{daSilva23} and broad Pfund lines detected in the NIRSpec spectra presented here.}
\tablebib{(a)~\citet{Tully88}; (b) \citet{Tonry01}; (c) \citet{Yuan21}; (d) \citet{Thim04}; (e) \citet{Bentz19}; (f) \citet{Koshida17}. $^{\S}$ With adjustments from \citet{Blakeslee01}.}
\label{tab1} 
\end{table*}

\section{Targets and JWST observations}
\label{observations}

  \begin{figure*}[!h]
  \centering
 {\par\includegraphics[width=1.0\columnwidth]{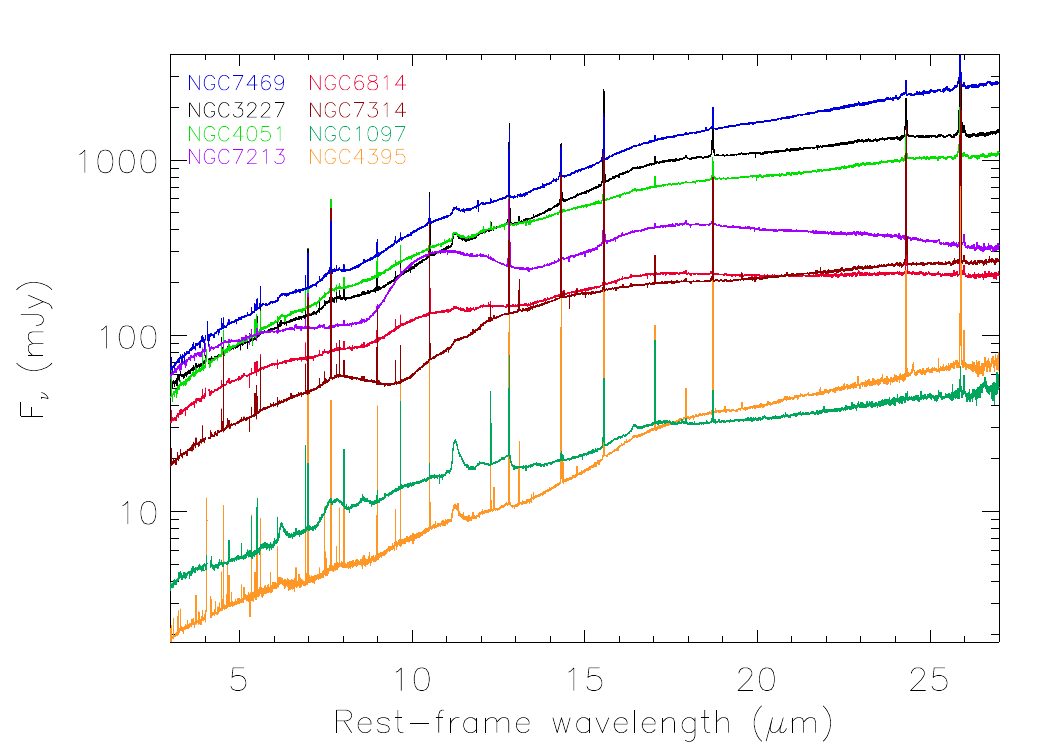}
  \includegraphics[width=1.0\columnwidth]{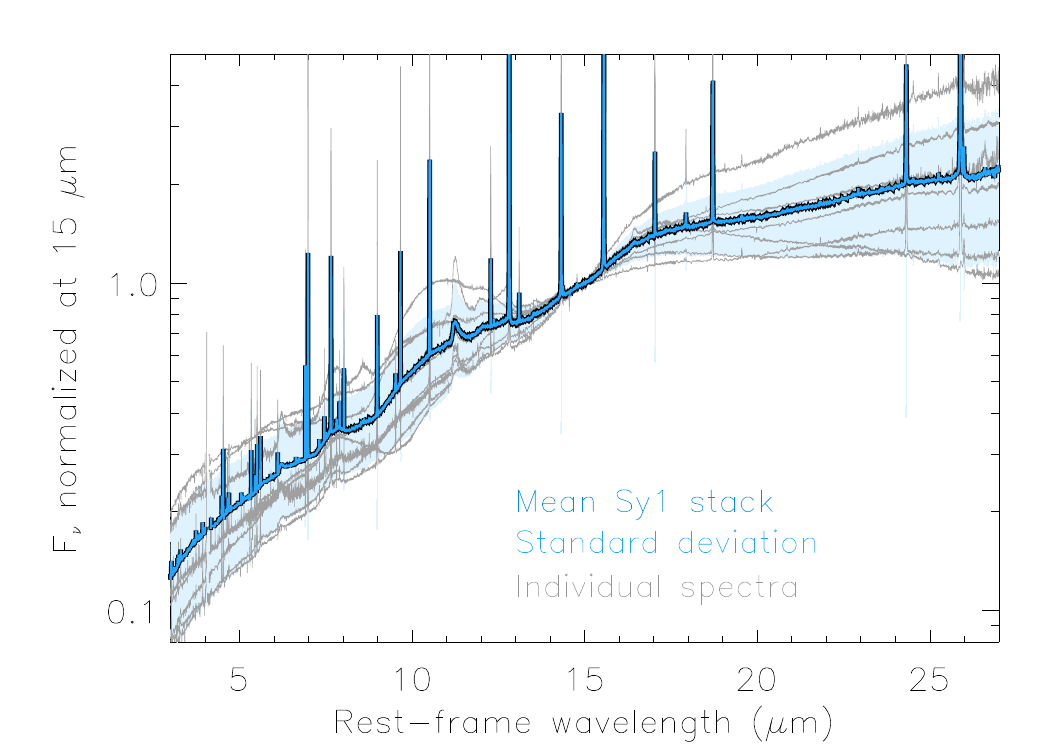}\par}
  \caption{Left panel: Infrared nuclear spectra of the eight Seyfert 1 (Sy1) galaxies, obtained from NIRSpec and MIRI observations. Right panel: Individual Sy1 spectra normalized at 15\,$\mu$m (grey lines), resulting mean stack (blue line), and standard deviation (light blue area).}
              \label{fig1}
    \end{figure*}

We selected the eight Seyfert 1 (Sy1) galaxies that had publicly available spectra from the JWST instruments MIRI and NIRSpec by the end of 2025. Most of the archival NIRSpec observations of nearby galaxies only cover the $\sim$3-5 $\mu$m range because it includes features such as the 3.3 and 3.4 \micron~policyclic aromatic hydrocarbon bands (PAHs) and because it is difficult to observe it with ground-based observatories due to atmospheric transmission. For the sake of homogeneity, we downloaded NIRSpec and MIRI integral field unit observations covering the range $\sim$3-28 $\mu$m. This includes the whole wavelength range covered by MIRI/MRS ($\sim$5-28 $\mu$m; R$\sim$1300-3700; $\rm \sigma_{inst}\simeq$35-100 km~s$^{-1}$;  \citealt{2021A&A...656A..57L,Argyriou23}) and the 2.87-5.27 $\mu$m range provided by NIRSpec using the grating and filter combination G395H/F290LP (R$\sim$2700; $\rm \sigma_{inst}\simeq$50 km~s$^{-1}$; \citealt{Jakobsen2022,Boker2023}). Most of the data come from the Galaxy Activity, Torus, and Outflow Survey (GATOS) JWST General Observer (GO) programs 3535 (PI: I. García-Bernete), 4972 (PI: L. Zhang), and 7195 (PI: A. Alonso-Herrero), except in the case of NGC\,4395 (GO program 2016; PI: A. Seth) and NGC\,7469 (Early Release Science Program 1328; PI: L. Armus). 

The data were reduced using the same version of the JWST Science Calibration Pipeline as in \citet{GarciaBernete22} and \citet{Pereira22}. Additional correction steps to correct hot and cold pixels were done, outlined in \citet{GarciaBernete24b} for the case of MIRI/MRS data and in \citet{Pereira2024} for NIRSpec.

The MIRI nuclear spectra were extracted as point source from all the MRS sub-channels, using aperture sizes ranging from 0.3 to 0.8\arcsec~with increasing wavelength. The spectra of the 12 bands were then stitched together using small scaling factors to make them coincide in the overlapping regions (see \citealt{Ramos25} for further details). We then stitched the NIRSpec nuclear spectra, extracted in an aperture size of 0.1\arcsec, to the nuclear MIRI spectra, taking advantage of their overlapping region at $\sim$5 $\mu$m. The multiplicative factors that we applied to the NIRSpec spectra range from 0.94 to 1.07. The resulting nuclear infrared spectra of the eight Sy1 galaxies are shown in Fig. \ref{fig1}.

The basic properties of the sample, including line-of-sight column density (N$_{\rm H}$), X-ray and bolometric luminosity, BH mass, and Eddington ratio are shown in Table \ref{tab1}. They have luminosity distances below 60 Mpc and relatively low galaxy inclinations (see Table \ref{tab1}) that allow us to isolate the central, AGN-dominated regions, minimizing the host galaxy contribution. Considering the angular resolution of NIRSpec ($\sim$0.1$\arcsec$~at 3 $\mu$m) and MIRI ($\sim$0.3-0.8$\arcsec$~across the whole range of 5-28 $\mu$m), the nuclear spectra probe the central $\sim$2-17 pc of the closest target, NGC\,4395, and $\sim$28-225 pc for the most distant one, NGC\,7469 (see Table \ref{tab1}).

\section{Results}

NGC\,1097 and NGC\,4395 are the dimmest nuclei in the sample, and NGC\,7469 is the brightest (see Fig. \ref{fig1}). The spectra show tens of emission lines and several PAH bands in the case of galaxies with nuclear star formation such as NGC\,1097 and NGC\,7469. The 9.7 $\mu$m silicate feature varies from weak absorption (NGC\,7314) to strong emission (NGC\,7213), although in the majority of the galaxies is either in weak emission or flat. This was already seen from ground-based mid-infrared spectra of Sy1 galaxies \citep{alonso16}, and it is consistent with clumpy (i.e. non-smooth) and two-phase torus model predictions \citep{Ramos09b,Ramos11b,Honig10,Alonso11,GarciaBernete19,GarciaBernete22,Gonzalez2019,Gonzalez2023}.

All the targets were previously observed with the InfraRed Spectrograph (IRS) of the Spitzer Space Telescope, whose spectra cover a much larger fraction of the galaxies than JWST (the central $\sim$3.6-10.7\arcsec~in the case of IRS versus $\sim$0.1-0.8\arcsec~for NIRSpec+MIRI). In Fig. \ref{figA1} we show a comparison between the JWST and Spitzer spectra of the eight Sy1 galaxies. From this comparison we can see that the targets with AGN-dominated infrared spectra, which have little or no PAH emission (NGC\,4395, NGC\,6814, and NGC\,7213), show similar JWST and Spitzer spectra. On the other hand, NGC\,1097, NGC\,3227, NGC\,4051, and NGC\,7469 show strong PAH bands in their Spitzer/IRS spectra that are either not present or much weaker in the JWST spectra. Finally, the Spitzer/IRS spectrum of NGC\,7314 shows deeper silicate absorption than the JWST/MIRI spectrum, indicating a host-galaxy contribution of cold dust that the JWST spectrum does not include. This qualitative comparison shows that even in the case of Sy1 galaxies, the nuclear JWST/NIRSpec+MIRI spectra is significantly different from those observed with Spitzer/IRS, except in the case of targets with AGN-dominated infrared spectra.

\subsection{JWST average stacked spectrum of Seyfert 1 galaxies}
\label{stack}

%One of the goals of this work is providing the community with a subarcsecond resolution infrared spectrum of Sy1 galaxies. 
%The JWST spectra shown in Fig. \ref{fig1} probe the central $\lesssim$200 pc of the galaxies (see Section \ref{observations}). 
To produce the Sy1 stack, we first interpolated the rest-frame individual spectra shown in the left panel of Fig. \ref{fig1} to a common wavelength grid. We adopted the wavelength sampling of the closest target, NGC\,4395, which sets the starting wavelength of the new grid (x$_1$=2.87 $\mu$m) up to an end wavelength of x$_2$=28.2 $\mu$m, defined by the most distant galaxy, NGC\,7469. 

Prior to stacking, the regrid spectra were normalized to their corresponding mean flux density within a narrow window centered at 15 $\mu$m (14.98–15.01 $\mu$m). These flux densities at 15 $\mu$m range from 17 to 891 mJy, depending on the target, with a mean value of 348$\pm$317 mJy. The normalization ensures that all spectra contribute equally at the reference wavelength while preserving relative differences in spectral shape (see right panel of Fig. \ref{fig1}). We chose to normalize at 15\,$\mu$m because it provides a relatively clean measure of the mid-infrared continuum while minimizing contamination from strong spectral features and variations in spectral slope. At shorter wavelengths (e.g. $\sim$12\,$\mu$m), the continuum can be affected by the wings of the 9.7\,$\mu$m silicate feature and/or 11.3 $\mu$m PAH band, introducing object-dependent biases when scaling the spectra. At longer wavelengths (e.g. $\sim$20\,$\mu$m), the emission becomes increasingly sensitive to variations in the warm-to-cool dust ratio, leading to a larger dispersion in spectral slopes among sources. In contrast, the region around 15\,$\mu$m is featureless and retains high signal-to-noise, providing a more stable and representative anchor for the intrinsic AGN continuum. In any case, we verified that the variations introduced in the stack by normalizing at 12 and 20 $\mu$m are smaller than the standard deviation of the stack normalized at 15 $\mu$m (see the right panel of Fig. \ref{fig1}).

The resulting set of uniformly sampled and normalized spectra was then used to construct the Sy1 stack, computed as the mean of the individual spectra at each wavelength. The associated dispersion was quantified via the standard deviation. This procedure yields an average spectrum, shown in the right panel of Fig. \ref{fig1}, while providing a measure of the object-to-object variability. The stack is representative of the eight Sy1 galaxies here considered, but not necessarily of the local Sy1 population. Nevertheless, these Sy1s cover a relatively broad range of Eddington ratios, BH masses, X-ray and bolometric luminosities (see Table \ref{tab1}). %As described in Section \ref{observations}, the sample was built from all the Sy1s with available NIRSpec+MIRI spectra at the time this work was initiated. 
%An ascii file containing four columns which correspond to the rest-frame wavelength, mean and median stacked spectra, and standard deviation, can be downloaded from the \href{https://gatos-astro.com}{GATOS} webpage. 
Neither the stack nor the individual spectra were corrected from extinction. For Sy1 galaxies and intermediate types, typical $\rm A_V$ values measured from ratios of Balmer and Paschen narrow emission lines and infrared-to-optical colors are $\lesssim$1-2 mag \citep{Landt08,Schnorr2016,Prieto2021}. This would correspond to $\rm A_{15\mu m}\lesssim$0.02-0.04 mag using the extinction-law of \citet{Chiar06}.

\subsection{Torus models}
\label{torus}

\begin{figure}[!h]
  \centering
 \includegraphics[width=1.0\columnwidth]{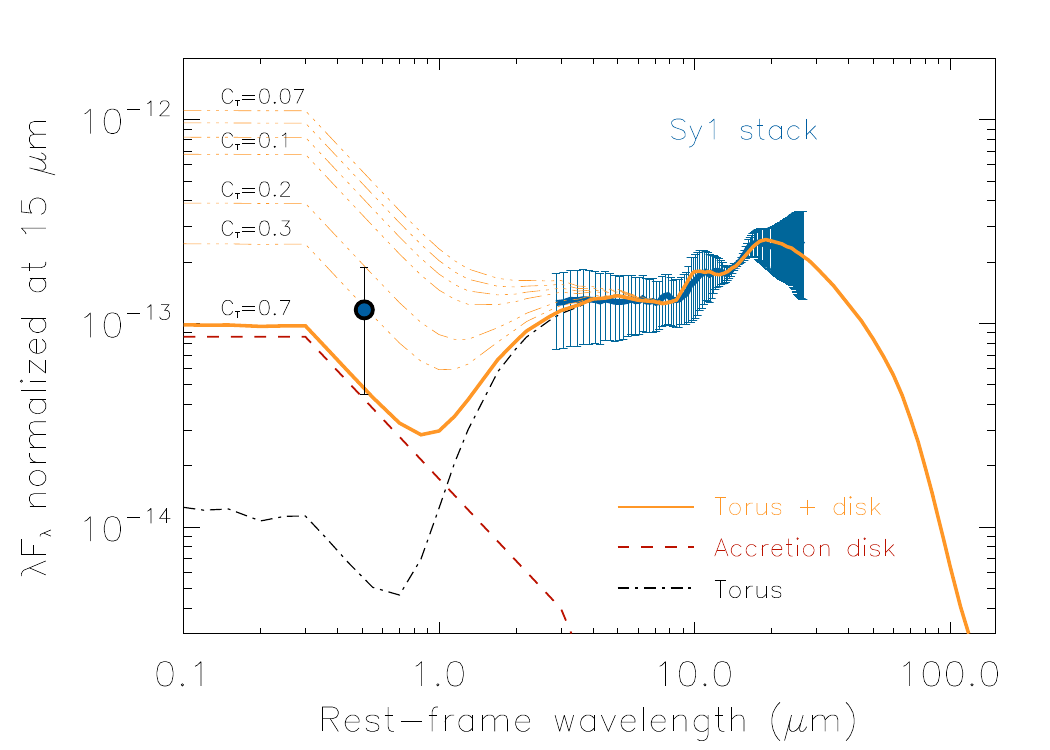}
  \caption{Median CAT3D-wind torus model (black dot-dashed line) fitted to the Sy1 stack shown in Fig. \ref{fig1} (blue solid line). The four-component power-law used as input spectrum to the CAT3D-wind models is the red dashed line. The orange lines are the sum of the median torus model and the accretion disc, scaling the disc to have energetic covering factors from $\rm C_T$=0.7 (solid line) to $\rm C_T$=0.3, 0.2, 0.1, 0.09, 0.08, and 0.07 (dot-dashed lines). The blue dot is the average flux at 5100 \AA~of the eight Sy1s, estimated from the $\rm \lambda F_{5100}^{H\alpha, avg}$ values reported in Table \ref{tab3}, normalized by the average flux at 15 $\mu$m ($\rm F_{\nu}$=347.5 mJy).}
%  and the dot-dashed orange lines are the same, but scaling up the accretion disc to have energetic covering factors of $\rm C_T$=0.2, 0.1, 0.09, 0.08, and 0.07, from bottom to top.} %The average flux at 5100 \AA~of the eight Sy1 galaxies, estimated from the $\rm \lambda F_{5100}^{H\alpha, avg}$ values reported in Table \ref{tab3}, and normalized by the average flux at 15 $\mu$m (348 mJy), is shown for comparison. The same torus models without any foreground extinction are also shown with lighter color.}
              \label{fig2}
    \end{figure}

To facilitate the comparison with different datasets including bluer and redder wavelengths than those covered by our Sy1 stack, we fitted it with \href{https://github.com/aasensio/bayesclumpy2}{\tt BayesClumpy2} \citep{Ramos26}. This new version of the Bayesian tool {\tt BayesClumpy} \citep{Asensio09,Asensio13} allows the user to fit photometry or spectroscopy data with the CLUMPY torus models of \citet{Nenkova08a,Nenkova08b} and the CAT3D-WIND models of \citet{Hoenig17}. We used an emission-line free, smoothed, and interpolated version of the Sy1 stack. We smoothed the spectrum using a Lee filter algorithm with a box size of 100, and then interpolated it to a grid of 120 wavelengths that successfully captures the spectral shape, including the silicate features (see blue solid line in Figs. \ref{fig2} and \ref{figC1}). %The 23 $\mu$m silicate feature, which has been detected in the nucleus of nearby obscured quasars with JWST \citep{Ramos25}, is not present in the spectra of the Sy1 galaxies studied here.

\begin{table*}[h!]
\centering
\footnotesize
  \caption{Properties of H$\alpha$ broad compiled from the literature.} 
\begin{tabular}{lcccccccccc}
\hline
\hline
ID   & FWHM & $\rm F_{\rm H\alpha}^{avg}\times10^{13}$ & $\rm F_{\rm H\alpha}\times10^{13}$ & $\rm F_{\rm H\alpha}^{max}/F_{\rm H\alpha}^{min}$ & $\rm log~L_{H\alpha}^{avg}$ & $\rm log~L_{H\alpha}$ & Refs. & $\rm H\alpha^b/H\beta^b$ & $\rm A_V^b$ & Refs. \\
  & (km~s$^{-1}$) & ($\rm erg~s^{-1}~cm^{-2}$) &  ($\rm erg~s^{-1}~cm^{-2}$) & & (erg s$^{-1}$) & (erg s$^{-1}$) & & & (mag) & \\
\hline
NGC1097 & 10000 &   1.045  &	 [0.306, 2.585]	 & 8.45 &   39.42  & [38.89, 39.81] &  a,b    & 2.8$\pm$0.1 & 0.0 & i \\
NGC3227 & 2900  &   18.74  &	 [13.49, 24.10]	 & 1.79 &   41.10  & [40.96, 41.21] &  c,d,e  & 3.1$\pm$0.7 & 0.0 & c,j \\
NGC4051 & 2000  &   11.76  &	 [11.22, 12.30]	 & 1.10 &   40.59  & [40.57, 40.61] &  d,e    & <2.8	   & 0.0 & d \\
NGC4395 & 1000  &   1.212  &	 [1.193, 1.230]	 & 1.03 &   38.43  & [38.42, 38.43] &  d,f    & <2.8	   & 0.0 & d  \\
NGC6814 & 3300  &   14.45  &	       \dots 	 & \dots &  40.91  & [40.85, 40.97] &  d      & \dots	   & 0.4$\pm$0.4 & k \\
NGC7213 & 2300$^*$& 13.90  &	 [12.39,17.23]	 & 1.39 &   40.91  & [40.86, 41.00] &  g      & \dots	   & 0.0$\pm$0.3 & l \\
NGC7314 & 1300  &   0.706  &	 [0.670, 0.741]	 & 1.11 &   39.45  & [39.43, 39.47] &  d,h    & 9.3$\pm$0.6 & 3.0$\pm$0.2 & h \\
NGC7469 & 2700  &   31.04  &	 [26.60, 35.48]	 & 1.33 &   42.10  & [42.03, 42.15] &  c,d    & <2.8 & 0.0 & c \\
\hline 
\end{tabular}
\tablefoot{The second column is the FWHM of the H$\alpha$ broad component from \citet{Mejia22}, except in the case of NGC\,1097, that is from \citet{Schimoia12}. The mean, minimum, and maximum broad line fluxes were calculated from the values reported in the works indicated in column seven. Column five lists the variation in flux from minimum to maximum. Columns six and seven list the broad line H$\alpha$ luminosities calculated from the fluxes reported in columns 3 and 4 and the distances from Table \ref{tab1}. Columns 8 and 9 indicate the Balmer decrement measured from the broad H$\alpha$ and H$\beta$ components and corresponding $\rm A_V$ assuming and intrinsic H$\alpha$/H$\beta$=3.06 \citep{Dong2008} and adopting the extinction law of \citet{Cardelli89} with $\rm R_V$=3.1. The $\rm A_V^b$ values of NGC\,6814 and NGC\,7213 were taken from \citet{Schnorr2016} and \citet{Caglar2020}, where several line ratios were used to compute the extinction towards the BLR. *NGC\,7213 also has broad double-peaked lines with a velocity separation of $\sim$9000 km~s$^{-1}$, whose fluxes, from \citet{Schimoia17}, are included in the broad H$\alpha$ flux.}
\tablebib{(a)~\citet{Schimoia12}; (b) \citet{Schimoia15}; (c) \citet{Landt08}; (d) \citet{Mejia22}; (e) \citet{Balmaverde14}; (f) \citet{Gu24}; (g) \citet{Schimoia17}; (h) \citet{daSilva23}; (i) \citet{Storchi97}; (j) \citet{Devereux2021}; (k) \citet{Schnorr2016}; (l) \citet{Caglar2020}.}
\label{tab2} 
\end{table*}

We ran the fits using nested sampling, uniform priors for all the parameters, and no foreground extinction. %, and accretion disk emission because it is a type-1 AGN stack. We also ran the fits without the accretion disk component (torus only) for comparison. %We also applied the extinction law of \citet{Chiar06} and a prior of $\rm A_V$=[0, 2] mag to account for some foreground extinction, unrelated to the torus. 
As flux error, we used the standard deviation of the stack (shaded area in the right panel of Fig. \ref{fig1}) plus 10\% of calibration uncertainty \citep{Ramos26}, summed in quadrature. 

Figure \ref{figC1} shows the fitted CLUMPY and CAT3D-wind models (those between the 16 and 84 percentiles of the posterior distribution) in orange. The black lines are the median model and corresponding residuals, which in the case of the CAT3D-wind model are smaller than 15\% across the whole wavelength range considered in the fit. The CAT3D-wind models better reproduce the shape of the mean Sy1 stack than the CLUMPY torus models, which produce larger residuals in the near-infrared (up to $\sim$50\% in the case of the median model). However, it is noteworthy that the residuals from the fit with the CLUMPY model at $\lambda$>20 $\mu$m are smaller than those from the fit with the CAT3D-wind model. Previous work on SED fitting of nearby AGN reported evidence for torus models including polar emission better accounting for the infrared emission of the majority of type-1 AGN than those without \citep{GonzalezMartin19,Gonzalez25,GarciaBernete22}. %The covering factor\footnote{%RA2026:$\rm C_{geo}=1-\int e^{-N_{LOS}(i)}d cos(i)$, with $\rm N_{LOS}(i) = N_0~f_{wd}~e^{-((90-i)-\theta_w)^2/2 \sigma^2} + N_0~e^{-(tan(90-i))^2/2 h^2
%See Table 3 in \citet{Ramos26} for a description of the torus parameters.} that corresponds to the fitted CAT3D-wind models shown in the left panel of Fig. \ref{figC1} is $\rm C_{geo}=0.8\pm0.1$. This is a relatively high covering factor for a type-1 AGN, due to the obscuration produced by both the disc and the wind, but since the viewing angle derived from the fit is 13$\pm^5_7$ deg (i.e., close to face-on), the silicate feature is seen in emission. 

%If we use the standard deviation as the error for the 5100 \AA~flux, the fitted torus models under-predict it (see black lines in Fig. \ref{fig2}). In order to have a realistic infrared-to-optical ratio, we tried two different approaches. The first one was reducing the error of the optical data point to 10\% to force the models to reproduce it (see brown line in \ref{fig2}). % and the foreground extinction (unrelated to the torus) is $\rm A_V =1.1\pm^{0.5}_{0.7}$ mag. The median model including and excluding the accretion disk component, which cover a rest-frame range from 0.01 to 36000 $\mu$m, are available to the community, so they can be easily compared with different datasets. 
Since we are working with type-1 AGN, we added the AGN input spectrum of the CAT3D-wind models (the four-component power-law described in Appendix A of \citealt{Honig2010b}), shown as a red dashed line in Fig. \ref{fig2}, to the median torus model (black dot-dashed line in Fig. \ref{fig2}). The resulting model SED, which accounts for both torus and accretion disc direct emission, is the orange solid line in Fig. \ref{fig2}. If we use Eq. 2 in \citet{Maiolino07} to derive the energetic covering factor ($\rm C_T$), but considering a factor of 9 instead of 7 to convert from L$_{\rm bol}$ to L$_{\rm 5100}$ \citep{Stern2012}, 
\begin{equation}
\label{eq1}
    \rm C_T = L_{dust} / L_{bol} \approx 0.3 \cdot \lambda L_{\lambda}(6.7~\mu m) / \lambda L_{\lambda}(5100~\AA)
\end{equation}

\noindent
we get $\rm C_T=0.7$. This energetic covering factor is often referred to as reprocessing efficiency (RE) of the torus or reprocessed fraction of the accretion disc emission \citep{Ramos11b,Stalevski2016}, and it shall not be mistaken with the geometrical covering factor derived from the torus model parameters ($\rm f_c$ or $\rm f_2$; see \citealt{Ramos17}). 
%, of 0.8$\pm$0.1 (see \citealt{Ramos26} for details). 

%Instead, if we use the average 5100 \AA~flux (see blue dot in Fig. \ref{fig2}), and we added it to the median torus model (black dot-dashed line in Fig. \ref{fig2}). 
%\noindent
%we get $\rm C_T=0.3$. The covering factor inferred from the CAT3D-wind model, $\rm C_{geo}$, is larger than the energetic covering factor because the former is purely geometrical, whereas the luminosity ratio is affected by anisotropic accretion-disk illumination and dust emission, and by the adopted bolometric corrections \citep{Stalevski2016}. 

Figure \ref{fig2} also shows the model SEDs resulting from scaling up the accretion disc emission to get $\rm C_T$=0.3, 0.2, 0.1, 0.09, 0.08, and 0.07 from Eq. \ref{eq1}. These model SEDs are used in Section \ref{LRDs}. Although the accretion disc and the torus SEDs should not be scaled independently if we seek for consistency between them, it can be done as an exercise to simulate AGN with different properties than those in the CAT3D-wind database, which were developed to reproduce local AGN. In any case, we checked that there are torus+disc models in the {\tt BayesClumpy2} database having the covering factors used here. %, and that the SEDs of these models at $\lambda \gtrsim$3-4 $\mu$m do not vary significantly for models with $\rm C_T=0.07-0.7$. Models with lower covering factors have slightly lower mid-infrared emission, but the SED shape remains practically identical. 

The model SED corresponding to $\rm C_T$=0.3 (orange solid line in Fig. \ref{fig2}) matches the average 5100 \AA~flux of the Sy1 galaxies (blue dot in Fig. \ref{fig2}). This average 5100 \AA~flux and its standard deviation was calculated from the $\rm \lambda F_{5100}^{H\alpha,avg}$ fluxes reported in Table \ref{tab3} and then %converted them to $\rm F_{\nu}$ and $\rm \nu F_{\nu}$. We obtain $\rm F_{\nu}$=6.9$\pm$4.2 mJy, which we then 
divided by the average flux of the Sy1 stack at 15 $\mu$m, the wavelength of normalization ($\rm F_{\nu}$=347.5 mJy).

%we The second approach was to take the CAT3D-wind models fits with and without accretion disc component (solid and dot-dashed black lines in Fig. \ref{fig2}). We then subtracted one from another, %and without forcing the models to reproduce the optical data point (i.e., using the standard deviation as the error), and scaled the accretion disk component to the 5100 \AA~flux (solid orange line in Fig. \ref{fig2}). By doing this we can reproduce the optical data point, and also account for its standard deviation (dashed orange lines in Fig. \ref{fig2}). The energetic covering factors of these scaled models are $\rm C_T$=0.2, 0.3, and 0.8. 

\begin{table*}[h!]
\centering
\footnotesize
  \caption{Continuum properties derived from H$\alpha$ broad fluxes and from single-epoch optical spectra.} 
\begin{tabular}{lccccccc}
\hline
\hline
ID &  $\rm \lambda F_{\rm 5100}^{H\alpha,avg}\times 10^{12}$  & $\rm log~\lambda L_{\rm 5100}^{H\alpha,avg}$ & $\rm log~\lambda L_{\rm 5100}^{H\alpha}$ & $\rm \lambda F_{ 5100}\times 10^{12}$ & $\rm log~\lambda L_{\rm 5100}$ & $\rm \lambda F_{\rm 6200}\times 10^{12}$  & $\rm log~\lambda L_{\rm 6200}$ \\
  & (erg s$^{-1}$ cm$^{-2}$) & (erg s$^{-1}$) &  (erg s$^{-1}$)  & (erg s$^{-1}$ cm$^{-2}$) & (erg s$^{-1}$) &  (erg s$^{-1}$ cm$^{-2}$) &  (erg s$^{-1}$) \\
\hline
NGC1097 & 5.597 & 41.15   &     [40.69, 41.49]  &  \dots   & \dots   &  \dots	&     \dots  \\
NGC3227 & 59.30 & 42.60   &     [42.48, 42.69]  &  63.10   & 42.63   &  48.98	&     42.52  \\
NGC4051 & 43.68 & 42.16   &     [42.14, 42.18]  &  53.70   & 42.25   &  85.11	&     42.45  \\
NGC4395 & 8.855 & 40.29   &     [40.29, 40.30]  &  2.754   & 39.78   &  3.090	&     39.83  \\
%NGC6814 & 48.55 & 42.44   &     [42.28, 42.59]  &  19.50   & 42.04   &  32.36	&     42.26  \\
NGC6814 & 62.55 & 42.54   &     [42.39, 42.70]  &  19.50   & 42.04   &  32.36	&     42.26  \\
NGC7213 & 46.73 & 42.43   &     [42.39, 42.51]  &  8.318   & 41.68   &  85.11	&     42.69  \\
NGC7314 & 25.04 & 42.00   &     [41.98, 42.02]  &  \dots   & \dots   &  1.122	&     40.65  \\
NGC7469 & 71.89 & 43.46   &     [43.40, 43.51]  &  107.2   & 43.63   &  104.7	&     43.62  \\
\hline 
\end{tabular}
\tablefoot{Columns 2-4 are the continuum flux, luminosity, and luminosity range at 5100 \AA~derived from the extinction-corrected broad H$\alpha$ fluxes reported in Table \ref{tab2} and using Eq. 1 in \citet{Greene2005}. Columns 5-8 are the single-epoch continuum fluxes and luminosities at 5100 and 6200 \AA~from \citet{Mejia22}. Luminosities were calculated using the distances reported in Table \ref{tab1}.}
\label{tab3} 
\end{table*}

\subsection{Connecting the BLR and optical continuum emission of Sy1 galaxies to their infrared spectra}
\label{normalizations}

%As described in Section \ref{intro}, one of the peculiar properties of z$\gtrsim$4 AGN, specially of LRDs, is their apparent infrared weakness \citep{Casey25,Chen2025,Setton2025,Xiao25}, as constrained from ALMA observations. 

  \begin{figure*}[!h]
  \centering
 {\par\includegraphics[width=1.0\columnwidth]{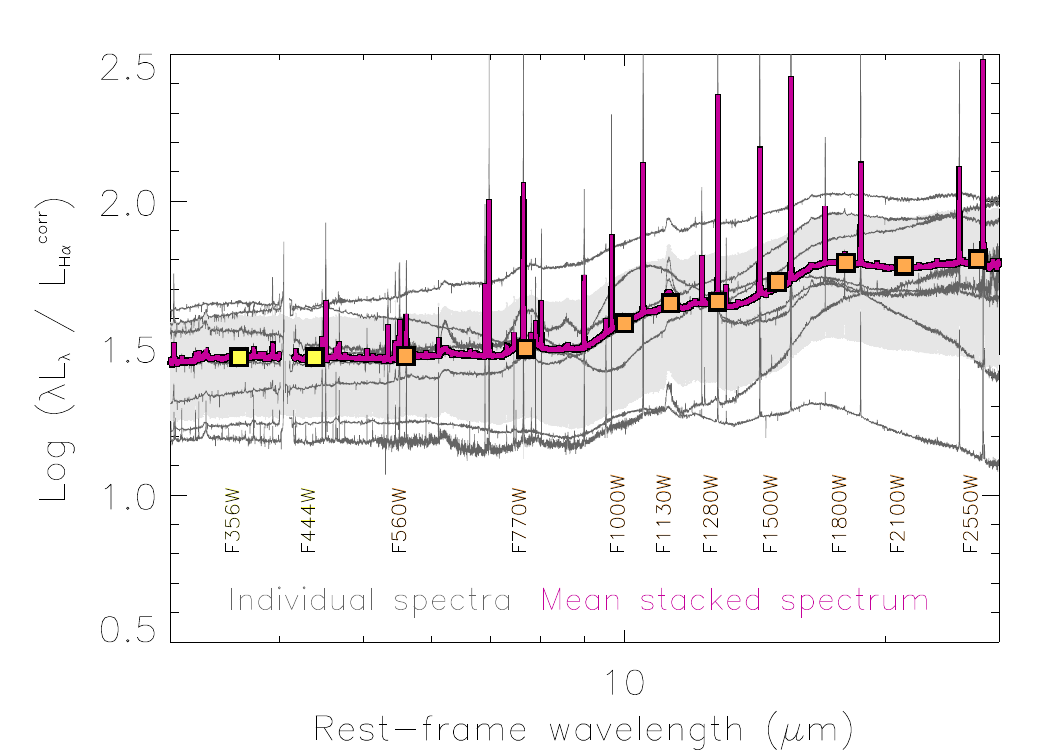}
  \includegraphics[width=1.0\columnwidth]{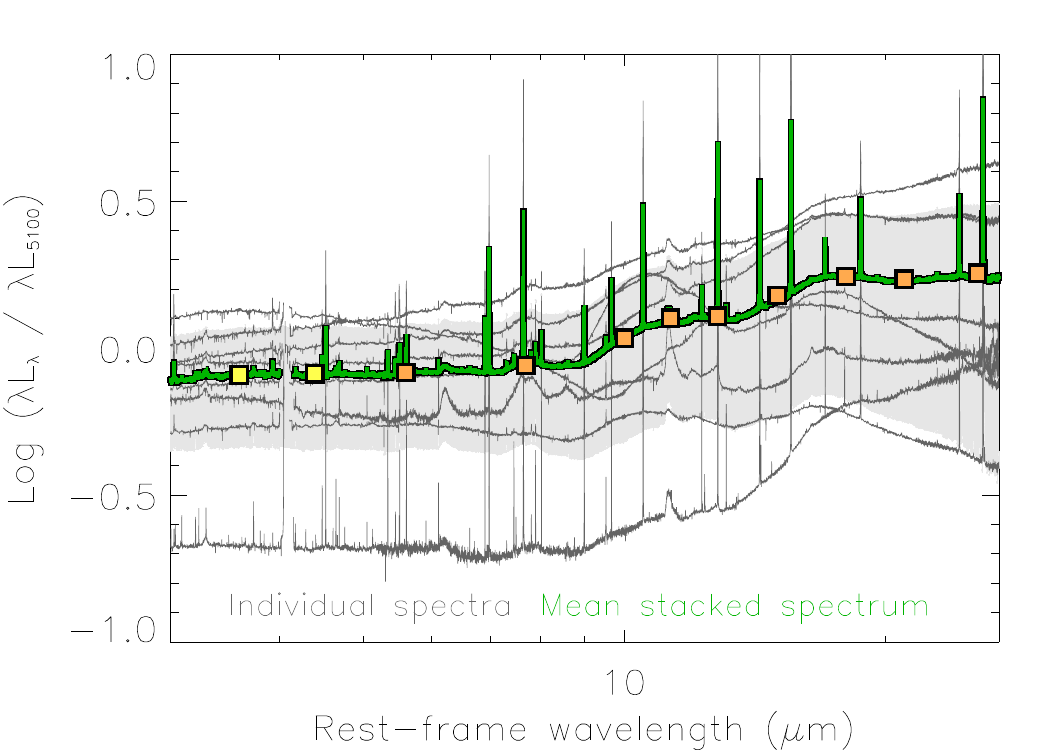}\par}
  \caption{Rest-frame spectra of the Sy1 galaxies (dark grey lines), normalized to their corresponding extinction-corrected broad H$\alpha$ luminosities (left panel) and 5100 \AA~luminosities derived from them  (right panel). Their corresponding mean stacks are shown in pink and green, respectively, with the grey shaded region indicating the standard deviation around the mean. Squares correspond to the average fluxes of the stacked spectrum measured in the wide MIRI filters (orange) and in the two NIRCAM wide filters F356W and F444W (yellow).}
              \label{fig3}
    \end{figure*}

Here we attempt to assess whether the rest-frame infrared emission of LRDs is consistent with predictions derived from our local Sy1 galaxies. In the literature, expectations on the infrared emission are set by the chosen normalization assumed from the optical. The issue with using the rest-frame optical continuum to predict the infrared at high-z is the uncertainty on whether that continuum is dominated either by the AGN or by the stellar component. For this reason, in principle it would be better to use the broad H$\alpha$ luminosity ($\rm L_{H\alpha}^b$) as normalizing quantity. The caveat then is that the large EWs measured for H$\alpha$ in distant AGN (see \citealt{Greene2026} and references therein) could imply that the covering factor of the BLR is higher than in local AGN \citep{Maiolino25,Yanagisawa2026} and/or that the H$\alpha$ emission is boosted by processes such as collisional excitation \citep{Brazzini26}. As a consequence, our local conversion from $\rm L_{H\alpha}^b$ to infrared luminosities might need to be re-scaled by a certain factor ($\sim$3-10; \citealt{Brazzini26}).

%We note, however, that from our fit with torus models, we obtain a covering factor of C$_T$=0.68$\pm$0.15, which is 

%In the Red Rosetta the amount of mid-IR radiation is a factor of 3–6 lower than expected from standard type I AGN templates, but this could be due to ionizing part of the UV radiation being filtered by the neutral absorber, and/or could also be associated with the lower metallicity (hence lower dust-to- gas ratio) in these systems.

An extra source of complication affecting optical normalizations for local type-1 AGN, both using the continuum or emission line luminosities, is variability. In the case of Sy1 galaxies, this variability usually ranges from weeks to months and it can be significant \citep{Peterson1998,Schimoia12,Schimoia15}. To account for this, we compiled broad H$\alpha$ fluxes for the eight Sy1 galaxies from the literature and calculated average values and ranges, which we report in Table \ref{tab2}. In some cases, these flux measurements span several years (e.g., NGC\,3227 and NGC\,4051). All the galaxies but NGC\,1097 have broad H$\alpha$ fluxes reported by \citet{Mejia22}, together with continuum measurements at 5100 and 6200 \AA~(see Table \ref{tab3}). According to the broad H$\alpha$ fluxes shown in Table \ref{tab2}, NGC\,1097 is the most variable source, with a 8.5 variation between maximum and minimum flux. The rest of the sources show variations ranging from 1.03 to 1.8\footnote{For NGC\,7469 there are also continuum and emission line fluxes measured from spectra taken between 1996 and 2015 from \citet{Shapovalova2017}, but we did not include their average H$\alpha$ flux of 37.3$\times10^{-13}$ $\rm erg~s^{-1}~cm^{-2}$ in Table \ref{tab2} because it is total instead of broad line flux.}. For NGC\,6814 we found only one measurement of the broad H$\alpha$ flux, and thus we allowed a 1.3 factor of variation, based on the average of all the Sy1 galaxies except the extreme case of NGC\,1097. The range of broad H$\alpha$ luminosities that we are probing with our Sy1 galaxies, uncorrected from extinction, is 38.4--42.2 erg~s$^{-1}$, with a mean value of 40.4$\pm$1.2 erg~s$^{-1}$. 

In Table \ref{tab2} we report the extinction measured from broad recombination line ratios ($\rm A_V^b$) reported in the literature. For all of the galaxies in the sample, $\rm A_V^b$<0.5 mag, in good agreement with the median values reported for BASS broad-line AGN \citep{Mejia22}, of 0.2 mag for galaxies with $\rm log~N_H\leq20~cm^{-3}$ and 0.6 mag for those with $\rm log~N_H>20~cm^{-3}$. The exception is NGC\,7314, for which \citet{daSilva23} reported A$_V^b$ values of 3.0 mag from the Balmer decrement of the broad components. This is related with the Seyfert 1.8/1.9 classification of the galaxy, which has changed to type 1/1.5 depending on when the observations were taken (see \citealt{daSilva23} and references therein). Therefore, we corrected its broad H$\alpha$ flux from extinction using the extinction law of \citet{Cardelli89} and R$_V$=3.1. We did the same for NGC\,6814, for which \citet{Schnorr2016} reported an $\rm A_V^b$=0.4$\pm$0.4 mag. Using the extinction-corrected $\rm L_{H\alpha}^{avg}$ values we get the same range of broad H$\alpha$ luminosities, and a mean value of 40.5$\pm$1.1 erg~s$^{-1}$. 

We then normalized the JWST spectra of the individual Sy1 galaxies, shown in Fig. \ref{fig1}, by their corresponding broad H$\alpha$ luminosities (using $\rm L_{H\alpha}^{avg}$ from Table \ref{tab2}, corrected from $\rm A_V^b$). The resulting spectra are shown in the left panel of Figs. \ref{fig3} and \ref{figB1}. In logarithmic scale, the normalized spectra vary between $\sim$1.1 and 2, with NGC\,4051 showing the highest values and NGC\,6814 the lowest. We also used the continuum luminosities at 5100 \AA~to normalize the spectra, which are shown in Figs. \ref{fig3} and \ref{figB1}. These continuum luminosities were calculated from $\rm L_{H\alpha}^{avg}$, corrected from extinction in the cases of NGC\,6814 and NGC\,7314, using Eq. 1 in \citet{Greene2005}. They are reported in Table \ref{tab3} and they range between 40.3 and 43.5 erg~s$^{-1}$ (mean value of 42.1$\pm$1.0 erg~s$^{-1}$). From the right panel of Figs. \ref{fig3} and \ref{figB1} it can be seen that the normalized spectra range from -0.7 to 0.6 in logarithmic scale, with NGC\,7469 showing the highest values and NGC\,4395 the lowest. The difference between the two normalizations shown in Fig. \ref{fig3} is of $\sim$1.6 dex ($\rm \lambda L_{5100}\simeq 40 \times L_{H\alpha}$), which comes from Eq. 1 in \citet{Greene2005}.

We note that although the conversion from $\rm L_{H\alpha}$ to $\rm L_{\rm 5100}$ might not be adequate for high-redshift AGN due to high H$\alpha$ EWs \citep{Greene2026}, we can safely use it for our local Sy1s. Indeed, for all the galaxies but NGC\,1097 and NGC\,7314 there are single-epoch 5100 \AA~continuum luminosities reported by \citet{Mejia22}, shown in Table \ref{tab3}, and they are in reasonable good agreement with the $\rm L_{H\alpha}^{avg}$-derived measurements (mean scatter of $\sim$0.3 dex). The largest discrepancy is found for NGC\,7213 (0.75 dex). We prefer to use $\rm log~L_{\rm 5100}^{H\alpha,avg}$ instead of the single-epoch values because they account for optical variability, which in the case of some of the targets is important (see Table \ref{tab2}), and there are  measurements for all the targets (see Table \ref{tab3}). %In any case, in Appendix \ref{appendixB} we show the infrared spectra normalized to the 5100 and 6200 \AA~continuum fluxes reported by \citet{Mejia22}, both shown in Table \ref{tab3}. 

Using the normalized spectra shown in the left and right panels of Figs. \ref{fig3} and \ref{figB1}, we produced the mean stacks shown in Fig. \ref{fig3}, using the same method as in Section \ref{stack}. In the case of the $\rm L_{\rm 5100}$ normalization, NGC\,4395 is below the standard deviation of the mean stack, but including it or not does not make any significant difference to the latter. The mean normalized stacks can be easily multiplied by the H$\alpha$ and 5100 \AA~continuum luminosities measured for high-redshift AGN in order to predict their rest-frame $\sim$3-27 $\mu$m emission. In Table \ref{tab4} we also report average values measured in the JWST MIRI and NIRCAM wide filters from the normalized stacks shown in Fig. \ref{fig3}. In the case of the broad H$\alpha$ normalization, $\rm \lambda L_{\lambda}/L_{H\alpha}^{corr}\simeq$[30,60], from 3 to 27 $\mu$m, and in the case of the 5100 \AA~normalization, $\rm \lambda L_{\lambda}/\lambda L_{5100}\simeq$[0.8,1.8]. %We note that the mean stacks shown in Fig. \ref{fig3} are the same as the Sy1 stack shown in the left panel of Fig. \ref{fig2}, but using the H$\alpha$ and 5100 \AA~continuum luminosities as normalization quantities instead of the 15 $\mu$m flux.

The bolometric luminosities of high-redshift AGN, including LRDs (log $\rm L_{\rm bol}~(erg~s^{-1}) \sim$ 44.1-44.4 for the LRD stack of \citealt{Perez26} and see also \citealt{Maiolino24,Geris2026,Scholtz2026}) are not that different anymore from those of local Sy1 galaxies. The normalizations provided by NGC\,7469, the most luminous Sy1 in our sample (log $\rm L_{\rm bol}~(erg~s^{-1}) \sim$ 44.5), are fully consistent with those of the stacks shown in Fig. \ref{fig3}. Therefore, despite the relatively low  luminosities of the Sy1 galaxies used here, of log $\rm L_{\rm bol}~(erg~s^{-1})=$41.8-44.5, they can be used to predict the infrared emission of LRDs with luminosities of at least up to log $\rm L_{\rm bol}~(erg~s^{-1}) \sim$44.5.

%We excluded NGC\,7213 from the stacks because its infrared continuum shape, dominated by the strong silicate features in emission, had a strong influence on the shape of the mean stacked spectra. This was not the case for the mean stacked spectrum described in Section \ref{stack} and shown in Fig. \ref{fig2}, due to the normalization we were using (at 15 $\mu$m). 

%\begin{figure*}[!h]
%  \centering
% {\par\includegraphics[width=1.0\columnwidth]{fig6a-eps-converted-to.pdf}
%  \includegraphics[width=1.0\columnwidth]{fig6b-eps-converted-to.pdf}\par}
%  \caption{Same as Fig. \ref{fig4}, but covering a broader wavelength range.}
%              \label{fig5}
%    \end{figure*}

%{\color {red} 
%Nancy mentioned that we would need to explain why high-redshift people should use our template rather than QSO templates considering the low-luminosities of these AGN. Here is where we can mention the halpha luminosities of the same in Roberto's JADES paper (2024), that are closer to the range we are measuring here.}

%Using JWST spectra of eight local Sy1 galaxies with spectral coverage from 3 to 27 $\mu$m, we provide the community with mean Sy1 stacks, normalized at 15 $\mu$m, $\rm H\alpha^b$, and 5100 \AA~continuum (see right panel of Fig. \ref{fig1} and Fig. \ref{fig3}), and torus models with different scaling of the accretion disc, which correspond to different energetic covering factors (see Fig. \ref{fig2}). 

\subsection{Predicting the rest-frame infrared emission of LRDs from H$\alpha$ and optical continuum measurements.}
\label{LRDs_normalizations}

We now use the Sy1 stacks shown in Fig. \ref{fig3} to predict the infrared emission of the z>2 LRD stack from \citet{Perez26} and the SED of the local LRD known as LORD/The Egg at z=0.1 \citep{Ji2026,Lin2026} from their broad H$\alpha$ and 5100 \AA~luminosities.

%and also with the local LRD known as LORD/The Egg at z=0.1 \citep{Ji2026,Lin2026}. In Section \ref{bolometric} we discuss the bolometric corrections to be applied to the broad H$\alpha$ and 5100 \AA~luminosities of LRDs and Sy1s. Finally, in Section 
%\ref{LRDs_normalizations} we test the predictions on the infrared emission derived from the Sy1 normalized stacks shown in Fig. \ref{fig3} with the LRD stack from \citet{Perez26} and the LORD SED from \citet{Ji2026}, in $\lambda$F$_{\lambda}$ units. 

%In Section \ref{LRDs} we simply scaled the Sy1 template and corresponding torus model fits to the $\sim$3 $\mu$m flux of the LRD stacks shown in Fig. \ref{fig4}. However, 

%Since not always there are rest-frame infrared fluxes available for LRDs, it is important to test whether it is possible to predict those from their broad H$\alpha$ luminosity and the optical continuum luminosity. %With this idea in mind, in Fig. \ref{fig3} we presented stacks of the spectra of the eight Sy1 galaxies considered here, after they were normalized by their corresponding broad H$\alpha$ and 5100 \AA~luminosities. 

%Using the 5100 \AA~and H$\alpha$ fluxes measured from the LRD stack from \citet{Perez26}, reported in Table \ref{tab5}, and from the optical spectra of LORD, at z=0.1 \citep{Ji2026,Lin2026}, 
%which are representative of the population of LRDs at 2.3<z<9.3 \citep{Perez26}, 
%we can test the predictions on their infrared fluxes using the Sy1 stacks shown in Fig. \ref{fig3}. 

\subsubsection{The infrared emission of LRDs at 2.3<z<9.3}
\label{pablo}

\begin{figure*}[!h]
  \centering
 {\par\includegraphics[width=1.0\columnwidth]{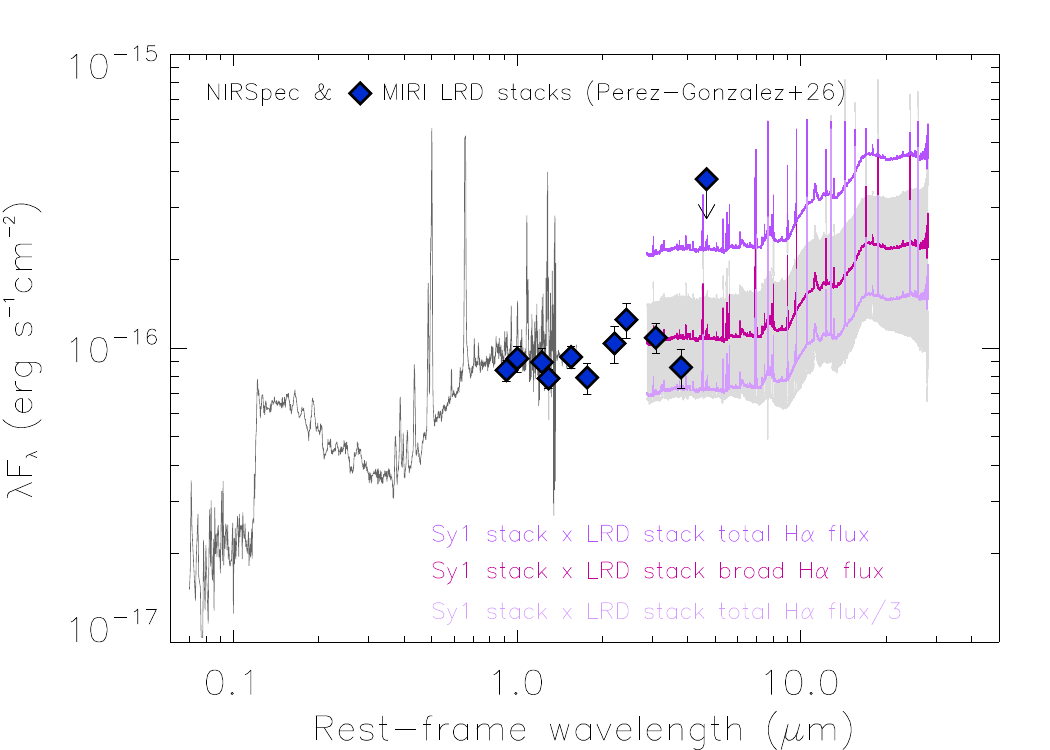}
  \includegraphics[width=1.0\columnwidth]{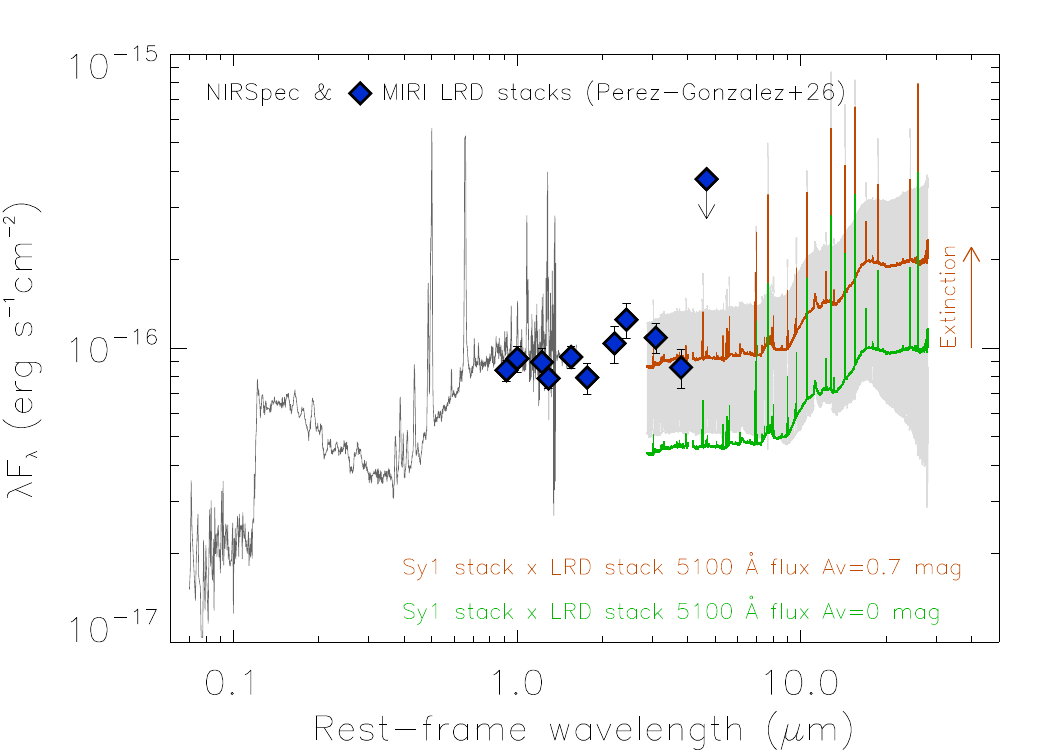}\par}
  \caption{Normalized Sy1 stacks from Fig. \ref{fig3}, shown as pink (purple) and green lines, multiplied by the broad (total) H$\alpha$ and 5100 \AA~fluxes measured from the LRD stack of \citet{Perez26}. The gray shaded area corresponds to the standard deviation around the average for the broad H$\alpha$ normalization (left panel) and for the extinction-corrected 5100 \AA~normalization (right panel). The lighter purple line is the total H$\alpha$ normalization divided by three. The brown solid line in the right panel is the normalized Sy1 stack, but multiplied by the 5100 \AA~flux corrected by an extinction of $\rm A_V$=0.7 mag to match the MIRI 3-4 $\mu$m LRD stacks. The NIRSpec Prism LRD stacked spectrum from \citet{Perez26}, in solid grey, and MIRI stacks (blue diamonds), in $\rm \lambda F_{\lambda}$ units, are shown in both panels.}
              \label{fig6}
    \end{figure*}

\begin{figure*}[!h]
  \centering
 {\par\includegraphics[width=1.0\columnwidth]{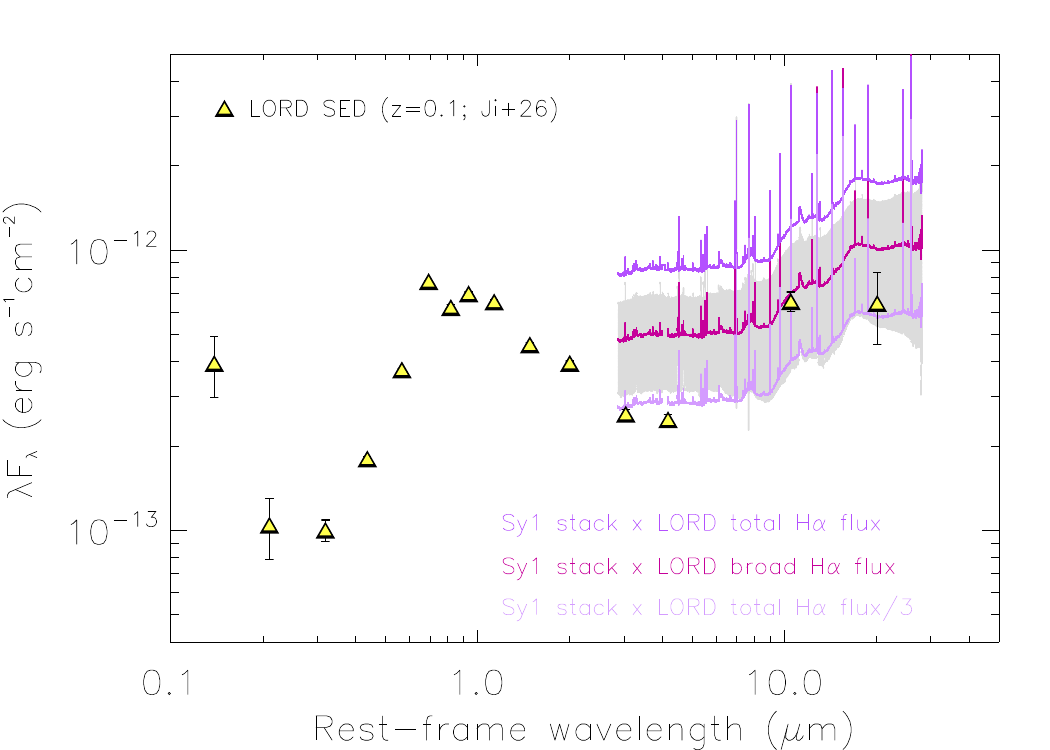}
  \includegraphics[width=1.0\columnwidth]{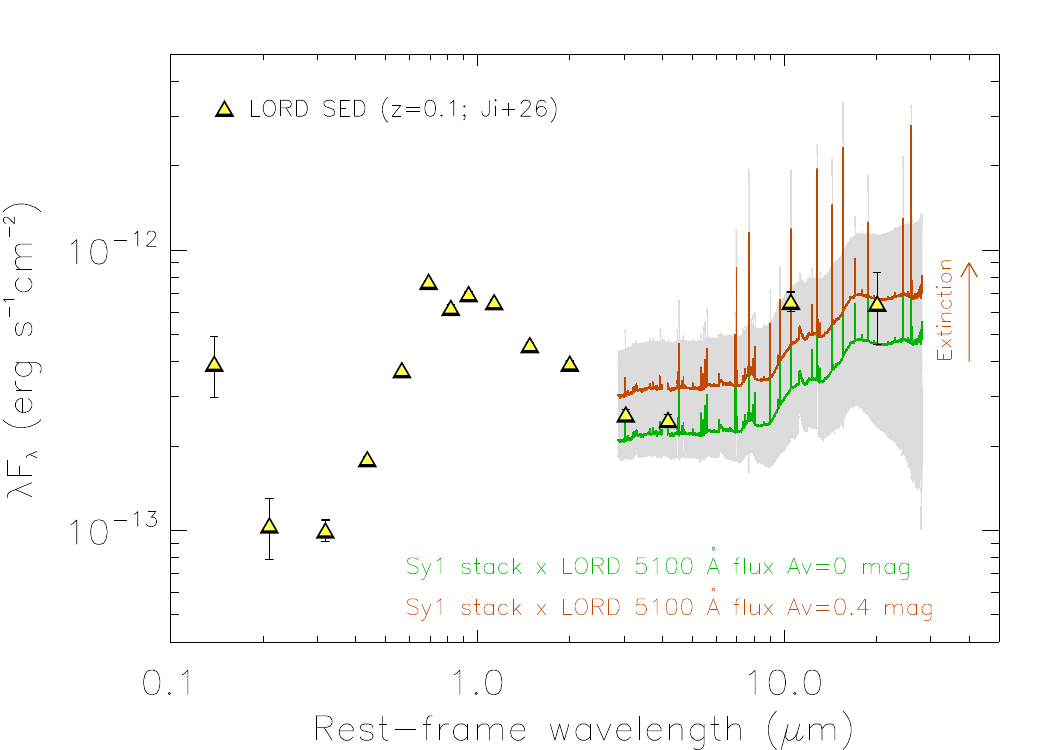}\par}
  \caption{Same as in Fig. \ref{fig6}, but for the local LRD (z=0.1) LORD \citep{Ji2026,Lin2026}.
  %Normalized Sy1 stacks from Fig. \ref{fig3}, shown as pink (purple) and green lines, multiplied by the broad (total=NB+BB) H$\alpha$ and 5100 \AA~fluxes measured from the Gemini spectrum of LORD from \citet{Ji2026}. The purple solid line in the left panel corresponds to the normalized Sy1 stack, but multiplied by the total broad H$\alpha$ flux (NB+BB components). 
  The brown solid line in the right panel is the normalized Sy1 stack, but multiplied by the 5100 \AA~flux of LORD corrected by the extinction of $\rm A_V$=0.4 mag reported by \citet{Ji2026} for the ISM/NLR of LORD. 
  %The gray shaded area corresponds to the standard deviation around the average for the BB H$\alpha$ normalization (left panel) and for the extinction-corrected 5100 \AA~normalization (right panel). 
  The LORD SED (yellow triangles) is shown in both panels.}
              \label{fig7}
    \end{figure*}

\citet{Perez26} stacked NIRSpec spectra of 249 LRDs at 2.3<z<9.3 obtained with the low resolution, R$\sim$30-330, double-pass prism (PRISM) and MIRI photometry of between 5 and 17 LRDs, depending on the filter (9, 12, 8, 10, and 5 LRDs for the five MIRI points at $\rm \lambda_{rest} \gtrsim 2~\mu m$). They reported a flattening of the MIRI SED at $\rm \lambda_{rest} < 2~\mu m$, in agreement with the NIRSpec Prism continuum, and a rising continuum at $\rm \lambda_{rest} \gtrsim 2~\mu m$, with a peak at 2-3 $\mu m$ (see Fig. \ref{fig6}). Here we consider the MIRI stack at $\rm 4.7~\mu m$ rest-frame as an upper limit, as it includes only five infrared-bright LRDs. \citet{Perez26} claim that the four MIRI stacks with rest-frame wavelengths between 2 and 3.8 $\mu m$ are consistent with the presence of hot dust at $\sim$1500 K. % and a rest-frame color of $\rm [1-3~\mu m]\sim 1.25$ mag. %This color is only slightly below the median one reported for MIRI F1800W-detected LRDs in \citet{Barro2026}, of $\rm [1-3~\mu m]\sim 1.5$ mag. 
From this NIRSpec LRD stack we measured a rest-frame 5100 \AA~flux of $\rm 5.70\times10^{-17}~erg~s^{-1}~cm^{-2}$ (see Table \ref{tab5}) in a narrow window between 5080 and 5120 \AA. %, after we converted the spectrum shown in the left panel of Fig. \ref{fig4} into $\lambda$F$_{\lambda}$ units. 
We used two Gaussian components of FWHM=3400 (broad) and 1050 km~s$^{-1}$ (intermediate) to reproduce the H$\alpha$ profile (see left panel of Fig. \ref{figD2} and Table \ref{tab5}), however, for our goals, the detailed functional form for fitting the H$\alpha$ profile is not critical.

%We measured the broad H$\alpha$ flux from the fit shown in the left panel of Fig. \ref{figD2}. We used two Gaussian components, of FWHM=3400 (broad) and 1050 km~s$^{-1}$ (intermediate), typical of high-redshift LRDs  \citealp{Matthee24,Greene2024,Brazzini26}), both resolved at the PRISM resolution (see Table \ref{tab5}). More complex profiles have been tested with higher spectral resolution data (see \citealt{Geris2026} and references therein), but at the PRISM resolution we rather keep the fit as simple as possible. We fitted two additional Gaussians to account for the [NII] emission, having the same FWHM as the ``narrow'' (intermediate, hereafter) H$\alpha$ component (see left panel of Fig. \ref{figD2}). HeI$\lambda$6678 \AA~is also detected in the LRD stack, overlapping with the red wing of H$\alpha$. The fluxes and luminosities of the broad and intermediate H$\alpha$ components are reported in Table \ref{tab5}.

In the left panel of Fig. \ref{fig6} we show the Sy1 stack from the left panel of Fig. \ref{fig3}, multiplied by the broad and total (broad+intermediate) H$\alpha$ flux of the LRD stack from \citet{Perez26}. The prediction from broad H$\alpha$ (pink line and shaded grey area in the left panel of Fig. \ref{fig6}) fully agrees with the $\sim$3-4 $\mu$m MIRI LRD stacks. %, showing that the broad H$\alpha$ flux nicely predicts the rest-frame near-infrared emission of LRDs at 2.3<z<9.3. 
It is noteworthy that the H$\alpha$ fluxes that we measure from the NIRSpec LRD stack do not include any extinction correction. If we correct the broad H$\alpha$ flux by different extinctions (e.g., the ones considered in \citealt{Perez26}, of $\rm A_V$=0.4, 0.7, and 1.6 mag), the predictions on the infrared emission of LRDs would be a factor 1.3-3.2 higher (see left panel of Fig. \ref{figD1}). Using the flux ratio H$\alpha$/H$\beta$ from Table \ref{tab5} %of 4.61$\pm$0.71 for the broad component, 
and the Small Magellanic Cloud (SMC) bar extinction curve of \citet{Gordon2003}, we get $\rm A_V$=1.1$\pm$0.4 mag, which would correspond to an infrared emission a factor of 2.2 higher than with an $\rm A_V$=0 mag. %This value of $\rm A_V$ is most likely an upper limit, based on the infrared predictions that we derive from the H$\alpha$.
We note that there is debate on the interpretation of the Balmer decrement of the broad components of LRDs, with models ranging from those entirely invoking density and radiative effects within the BLR (or surrounding cocoon) to those claiming significant dust reddening effects \citep[e.g.][]{DEugenio2025,Nikopoulos2026Balmer,Kiyota2026,Madau2026b,Madau2026Lord}. Here we do not enter into this debate and note that the inferred reddening is similar to the range reported by \cite{Perez26}.  %and, depending on the detailed interpretation, the H$\alpha$ flux correction would be up to a factor of $\sim$2.2. 
We will discuss more thoroughly the role of dust extinction in Section \ref{LRDs}.

In the left panel of Fig. \ref{fig6} we also show the Sy1 stack, but multiplied by the total (broad + intermediate) H$\alpha$ flux from Table \ref{tab5} (purple line), which is a factor of two higher than the broad H$\alpha$ flux. Given that the LRD stack was built from low resolution PRISM spectra, it is not clear whether the intermediate component traces a narrow (NLR/SF) component or it also contains part of the BLR emission. In the extreme case in which the total flux traces the full BLR emission, it would over-predict their infrared emission by a factor of $\sim$2. This would be consistent with a higher covering factor of the BLR in these systems \citep{Maiolino25,Brazzini2025,Yanagisawa2026}.
%However, the correction factor, of $\sim$1-3 if we consider the predictions from broad to total H$\alpha$, is lower than previously estimated for a high-covering factor BLR (e.g. $\sim$3-10; \citealt{Brazzini26}). %Alternatively, the infrared excess predicted from the total H$\alpha$ flux could be associated with the optical excess relative to the torus models fitted to the Sy1 stack, shown in Figs. \ref{fig4} and \ref{fig5}. 
However, in the latter scenario of BLR with high covering factor in LRDs, the H$\alpha$ scaling given by the Sy1 stack should be corrected by about a factor of 3 \citep{Brazzini26}. This is shown by the light purple line in the left panel of Fig.\ref{fig6}, which is also consistent with the observed data.

%In the left panel of Fig. \ref{fig6} we show the predicted infrared emission using the total H$\alpha$ flux divided by three (the total H$\alpha$ flux divided by two coincides with the broad H$\alpha$ normalization). 

In the right panel of Fig. \ref{fig6} we show the normalized Sy1 stack from the right panel of Fig. \ref{fig3}, multiplied by the 5100 \AA~flux measured from the LRD stack (see Table \ref{tab5}). The predicted infrared emission, shown as a green solid line, is below the MIRI LRD stacks, indicating that the 5100 \AA~flux might not be as good as a good predicting quantity as the broad H$\alpha$ flux. However, considering that H$\alpha$ (and particularly, H$\alpha$ broad) is less affected by extinction than the 5100 \AA~flux, we repeated the exercise of correcting the continuum flux by different extinction values. The results are shown in the right panel of Fig. \ref{figD1}. In the case of the 5100 \AA~flux, the predictions on the infrared emission of LRDs would be a factor 1.5-5 higher if we consider $\rm A_V$ values of 0.4, 0.7, and 1.6 mag. As can be seen from the right panel of Figs. \ref{fig6} and \ref{figD1}, an extinction-correction of $\rm A_V\sim0.7$ mag (a factor 2 higher in flux) nicely matches the MIRI 3-4 $\mu$m LRD stacks. %It it noteworthy that this level of reddening is fully consistent with the value of the foreground extinction of $\rm A_V=1.1\pm^{0.5}_{0.7}$ mag independently estimated from the tfoccorus model fit of the local Sy1s shown in Fig. \ref{fig2}.

\begin{figure*}[!h]
  \centering
 \includegraphics[width=1.4\columnwidth]{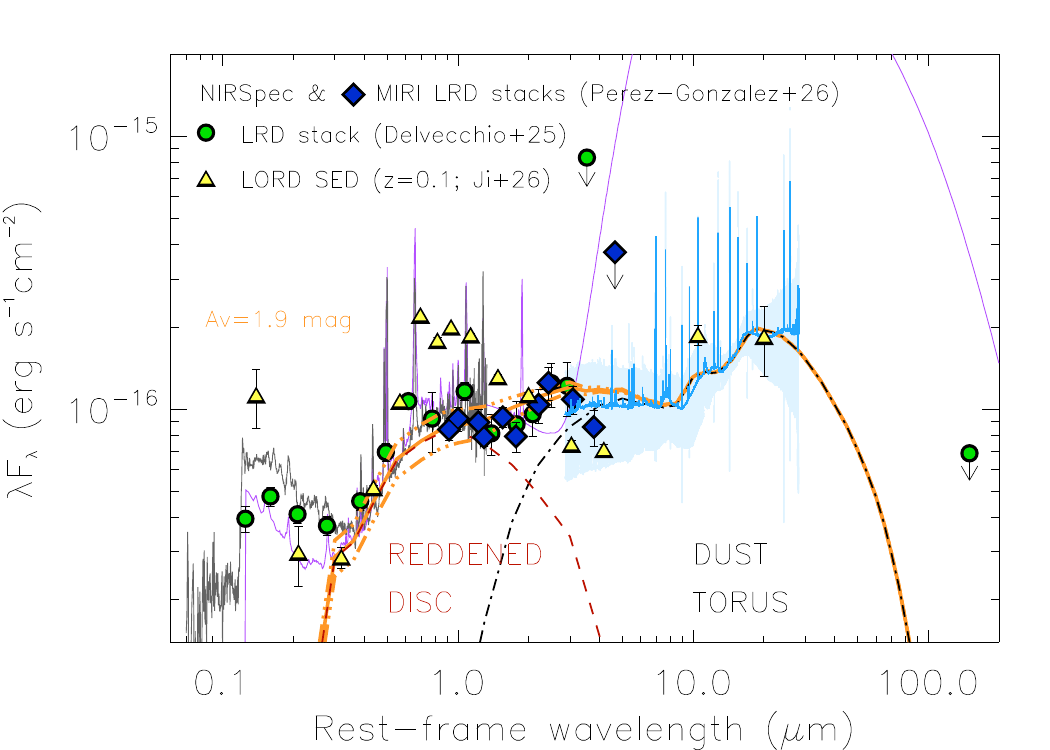}
  \caption{Sy1 stack and standard deviation (blue solid line and cyan shaded area) and median CAT3D-wind torus models (orange lines) with energetic covering factors of $\rm C_T$=0.07, 0.08, and 0.09 (see Fig. \ref{fig2}), reddened by $\rm A_V$=1.9 mag using the SMC bar extinction law. The accretion disc and torus components of the model with $\rm C_T$=0.08 are shown as red dashed and black dot-dashed lines, respectively, and the ratio between their integrated luminosities is $\rm L_{dust}/L_{disc}^{reddened}=2.8$ (see Table \ref{tab:lum}). The Sy1 stack without extinction (see Fig. \ref{fig2}) was scaled to the rest-frame 3 $\mu$m MIRI stack of \citet{Perez26}. The NIRSpec Prism median stacked spectrum is shown in solid grey, and the MIRI stacks as blue diamonds. We also show the photometric LRD stacks from \citet{Delvecchio2025} and \citet{Akins2025} as green dots and a solid light purple line. The stacked photometry from \citet{Delvecchio2025} was multiplied by a factor of 1.4 to match the other two stacks. The yellow triangles correspond to the SED of the local LRD LORD from \citet{Ji2026}, divided by 3500 to match the high-z LRD stacks.} %The Sy1 stack was multiplied by factors of 0.0054 and 0.0051 to match the $\sim$3-4 $\mu$m flux of the two LRD stacks. The torus models with and without accretion disk component include a foreground extinction (unrelated to the torus) of $\rm A_V$ =1.1 mag, derived from the fits. The same models without any extinction are also included for comparison.}
              \label{fig4}
    \end{figure*}

\subsubsection{The infrared emission of an LRD at z$\sim$0.1}
\label{lord}

In the case of the z$\gtrsim$4 LRD stacks, the longest wavelengths for which there are detections are $\sim$3-4 $\mu$m. For this reason, in Fig. \ref{fig7} we also show a comparison between the Sy1 stacks and the photometric SED of the local LRD (z=0.1) known as ``LORD'' \citep{Ji2026} or ``The Egg'' \citep{Lin2026}. This SED includes WISE fluxes from 3.4 to 22 $\mu$m that can be directly compared with the mid-infrared continuum of the Sy1 stack. Here we consider the broad (BB in \citealt{Ji2026}) and total (NB+BB, but excluding the narrow component of 90 km~s$^{-1}$) H$\alpha$ fluxes measured from the Gemini spectrum and reported in Table 1 from \citet{Ji2026}, and a 5100 \AA~flux of $\rm 2.722\times10^{-13}~erg~s^{-1}~cm^{-2}$, measured from the same spectrum. 

In the left panel of Fig. \ref{fig7} we show the Sy1 stack from the left panel of Fig. \ref{fig3}, multiplied by the broad and total H$\alpha$ fluxes of LORD. The broad and total H$\alpha$ fluxes overestimate the WISE W4 flux (22 $\mu$m) by factors of 1.6 and 2.7. The template's scatter around the broad H$\alpha$ normalization is consistent with the WISE W3 and W4 fluxes, although the W3 flux could be boosted by emission spectral features, such as the 9.7 $\mu$m silicate feature, the 11.3 and 12.5 $\mu$m PAH features, and/or emission lines such as [SIV]$\lambda$10.51 $\mu$m and [NeII]$\lambda$12.81 $\mu$m. Upcoming JWST/MIRI observations of LORD (GO proposal 12316; PI: F. Bauer) will reveal its mid-infrared spectrum. 

Similarly to the analysis of the high-z stack, in the left panel of Fig. \ref{fig7} we also show the normalization using the total H$\alpha$ flux divided by three, which matches the W1, W2, and W4 fluxes. As in the case of the high-z LRD stack of \citet{Perez26}, we are not considering extinction correction for the H$\alpha$ fluxes in the left panel of Fig. \ref{fig7}. 
%the H$\alpha$ normalizations shown in the left panel of Fig. \ref{fig7} would be indicating that the extinction of the BLR of LORD should be close to zero, as in local type-1 AGN. 
\citet{Ji2026} estimated $\rm A_V<5.1$ mag from the Balmer decrement, but they acknowledged that it can
be enhanced by collisional excitation \citep{Ilic2012}, resembling dust attenuation. 

In the right panel of Fig. \ref{fig7} we show the Sy1 stack from the right panel of Fig. \ref{fig3}, multiplied by the 5100 \AA~flux of LORD (green solid line). In this case, the continuum normalization matches the W1 and W2 fluxes and it underpredicts the W4 flux by a factor of 0.7. However, if we correct the 5100 \AA~flux from extinction using the value reported by \citet{Ji2026} for the ISM/NLR of LORD, of $\rm A_V=0.4$ mag, the continuum normalization accommodates all the WISE fluxes (brown solid line and gray area in Fig. \ref{fig7}).

%Thus, considering the analysis done in Sections \ref{pablo} and \ref{lord}, we conclude that the broad H$\alpha$ and 5100 \AA~fluxes of LRDs can be used to estimate their infrared emission using our local Sy1 stacks. In the case of the broad H$\alpha$ fluxes, extinction corrections would not be needed, based on the comparisons done here, and in the case of the optical continuum, small amounts of extinction $\rm A_V\simeq$0.4-0.7 mag are required.  

%Considering the EWs that we measure for the broad and total H$\alpha$ components of LRDs, of $\simeq$300-570 \AA~\citep{Perez26,Ji2026}, which fall within the range of EWs measured for local type-1 AGN (EW=[15,500] \AA~for $\rm log~L_{H\alpha}^b$=[40, 44] erg~s$^{-1}$; \citealt{Stern2012}), and the 5100 \AA~fluxes directly measured from rest-frame optical spectra, we conclude that the 5100 \AA~flux can be estimated either from the broad H$\alpha$ flux or from the total (broad+intermediate components) divided by a factor of $\sim$1.5-2. More predictions of high-redshift LRD stacks and local LRDs are required to test the predictions here discussed. 

Considering the comparisons presented in Sections \ref{pablo} and \ref{lord}, we find that the infrared emission of LRDs inferred from their H$\alpha$ luminosities (uncorrected from extinction) exceeds the actual infrared emission by a factor of $\sim$1--3, placing our estimates at the lower end of the factor of $\sim$3--10 predicted by \citet{Brazzini26} for a high covering factor BLR. However, this 1--3 factor would increase if $\rm A_V (BLR) >0$ mag. Overall, the interpretation of the results when normalizing by H$\alpha$ are not trivial, as the deviations can be interpreted either in terms of higher covering factor of the BLR, excess production of H$\alpha$ due to radiative and density effects in the BLR, or H$\alpha$ extinction -- the effects are degenerate at this level.

When we use the 5100~\AA\ flux, the predicted infrared emission is consistent with the actual values using a modest amount of extinction, of $\rm A_V\lesssim1$ mag (i.e. within a factor of $\sim$2). Despite the large galaxy areas probed by the LRD stack of \citet{Perez26} and, to a lesser extent, by the SED of LORD, our comparisons are consistent with an AGN-dominated optical continuum of LRDs. %If confirmed for more LRD stacks and individual sources, the 5100 \AA~luminosity can be used to predict their infrared emission using our local Sy1 stack. %For this reason, and considering that the broad H$\alpha$ flux can be affected by BLR covering factor, collisional extinction, and extinction, 
Therefore, here we favor the use of the 5100~\AA\ continuum as a predicting quantity. 
We discuss the AGN-origin of the optical emission of LRDs in Section \ref{discussion}.

\section{Discussion}
\label{discussion}

In Section \ref{LRDs} we compare the Sy1 stack and torus model fits shown in Fig. \ref{fig2} with different stacks of LRDs at z$\gtrsim$2 \citep{Delvecchio2025,Akins2025,Perez26}. 
In Section \ref{enbal} we briefly discuss the optical-infrared energy balance aspects of our model. Finally, in Section \ref{dust} we estimate the dust mass of the torus model resembling the optical and near-infrared SED of these LRD stacks and we compare it with different estimates published in the literature. % and with the SED of a local LRD (z=0.1; \citealt{Ji2026,Lin25}). 

\subsection{A reddened accretion disc and a low-reprocessing efficiency torus to explain the optical-to-infrared SED of LRDs}
\label{LRDs}

In Section \ref{LRDs_normalizations} we showed that the infrared emission of LRDs can be predicted from their 5100 \AA~continuum luminosity using the Sy1 stacks shown in Fig. \ref{fig3}. This approach has the advantage of being model-independent. However, the accretion disc+torus model that fits both the Sy1 stack and optical point (see Fig. \ref{fig2}) does not resemble the optical and infrared SED of LRDs, suggesting that their accretion disc and torus properties differ from those of Sy1s. We therefore attempt to reproduce the optical-to-infrared continuum emission of high-redshift LRDs by varying the torus covering factor (i.e., its reprocessing efficiency, which is the scaling of the accretion disc relative to the torus model) and the foreground extinction.

We scaled our Sy1 stack and fitted torus model (blue solid line and black dot-dashed line in Fig. 2) to the rest-frame 3 $\mu$m  MIRI stack from \citet{Perez26}. We then considered different scalings of the accretion disc component relative to the torus model (see Section \ref{torus} for details), which correspond to covering factors ranging from $\rm C_T$=0.7 to 0.07 (orange lines in Fig. \ref{fig2}). %As mentioned in Section \ref{torus}, the disc and torus components should not be  %We refer the reader to \citet{Madau2026b} for an example of energy-conserving low-covering factor dust models specifically developed to reproduce the SED of high-redshift LRDs.
%This is shown in Fig. \ref{fig4}. %The scaled Sy1 stack %, which we had to multiply by 0.0054, 
%nicely matches the spectral shape of the NIRSpec LRD stacked spectrum and MIRI stacks. 
%The torus model shows good agreement with the LRD MIRI stacks, and it is above the NIRSpec stack at $\lambda\lesssim$1 $\mu$m. The MIRI stack at 3.8 $\mu$m is below the model but still consistent with the standard deviation of the mean Sy1 stack. For comparison, we also include the torus models with $\rm C_T$=0.2 and 0.1 (dashed and dot-dashed line in Fig. \ref{fig4}). %This could be possibly indicating a flattening of the mid-infrared continuum at longer wavelengths, although its associated error is larger than those of the other MIRI stacks. 
To account for additional foreground extinction not included in the CAT3D-WIND models, we applied an extinction law, with $\rm A_V$ varying from 0.5 to 3 mag, to the Sy1 stack and model SEDs. We adopted the SMC bar extinction curve of \citet{Gordon2003}, which may be more appropriate for high-redshift LRDs \citep{Madau2026b}, but also for normal AGN and quasars at high-z \citep{Richards2003}. We scaled it % at $\lambda < 1.25\,\mu$m, 
to match the \citet{Chiar06} local ISM extinction curve at the K-band ($1.25\,\mu$m), which we used up to $27\,\mu$m, and then extrapolated it beyond this wavelength with a power law ($A_\lambda \propto \lambda^{-2}$).%, normalized to the Chiar \& Tielens (2006) Local ISM value of $A_{27\,\mu\mathrm{m}}/A_K = 0.452$. The extinction curve is expressed as $A_\lambda/A_K$ throughout.

We find that the torus model with C$_{\rm T}$=0.08$\pm$0.01 and an additional, foreground extinction of $\rm A_V$=1.9$\pm$0.1 mag, is the one that best captures the shape of the LRD stack for rest-frame $\lambda\gtrsim$0.3-0.4 $\mu$m, %, but it is the model with C$_{\rm T}$=0.2 and $\rm A_V$=1.5 mag that best captures the shape of the LRD stack at $\lambda\gtrsim$0.4 $\mu$m 
as illustrated by the orange solid line in Fig. \ref{fig4} (with the orange dot-dashed lines corresponding to the torus models with C$_{\rm T}$=0.07 and 0.09). This exercise shows that a reddened accretion disc 
%\footnote{We used the input AGN spectrum of the CAT3D-wind models, but the actual shape of the acretion disc spectrum is much less important for dust heating than its total luminosity \citep{Honig2010b}.} 
and a small covering factor torus can explain the shape of the optical and near-infrared continuum of high-redshift LRDs. The low torus covering factor or reprocessing efficiency (RE) indicates that the obscuring structures of LRDs differ from those of local type-1 AGNs, which typically exhibit intermediate values (\(\mathrm{L_{dust}/L_{bol}} \simeq 0.4\text{--}0.7\); \citealt{Ramos11b}).

%However, since the dust properties of high-redshift LRDs are necessarily different from those of local AGN (i.e., low-metallicity environments in the early universe), for which the CAT3D-wind models have been developed, here we attempted this exercise to simulate lower-covering factor tori. 

It is worth mentioning that the foreground extinction that we apply to the C$_{\rm T}$=0.08 model to match the optical-to-infrared shape of the LRD stack of \citet{Perez26}, of $\rm A_V=1.9$ mag, is higher than the value of 0.7 mag used in the right panel of Fig. \ref{fig6}. This is because the C$_{\rm T}$=0.08 model has an intrinsically brighter optical continuum than the 5100 \AA~average flux of the Sy1s (see Fig. \ref{fig2}), and thus requires a higher extinction to have the same optical-to-infrared ratio.  
%relative to the infrared than the C$_{\rm T}$=0.3 model, which is the one accommodating the 5100 \AA~average flux of the Sy1s and their infrared stack (see Fig. \ref{fig2}). 
%{\color {magenta} In other words, the optical to infrared ratio of the C$_{\rm T}$=0.08 with $\rm A_V=2.2$ mag and of the C$_{\rm T}$=0.3 with $\rm A_V=0.7$ mag models are similar, but they differ in optical shape. Considering this, does it still make sense to use the optical prediction for the infrared emission derived from Section \ref{LRDs_normalizations}? It has the advantage that is model independent, but we know that with that combination of covering factor (0.3) and extinction (Av<1 mag) we cannot reproduce the whole optical-to-infrared shape, just the optical/MIR ratio. I guess it is fine, but check this with Roberto.}

%lies below the LRD stack at $\lambda$<3.6 $\mu$m. On the other hand, the same model without foreground extinction, also shown in Fig. \ref{fig4}, lies below the rest-frame LRD stack at wavelengths between $\sim$0.5 and 3.6 $\mu$m, indicating the presence of a hotter component. Between $\sim$0.15 and 0.4 $\mu$m the torus model with $\rm A_V=0$ mag is above the LRD stack. %It is noteworthy that the torus models shown in Fig. \ref{fig4} include a 

Fig. \ref{fig4} also shows two photometric LRD stacks for comparison. The green dots correspond to the stack of 300 LRDs with redshifts of 2<z<11 from \citet{Delvecchio2025}, and the light purple solid line, to the COSMOS-Web  Maximal SED from \citet{Akins2025}. The latter is the maximal SED allowed by the photometry upper limits, which at $\rm \lambda_{rest} > 3~\mu m$ are from Spitzer, Herschel, SCUBA-2, and ALMA, and they include 434 LRDs at 5<z<9 (except in the case of the ALMA stacks, which include 30 and 128 LRDs at 1.2 and 2 mm; see Fig. 13 in \citealt{Akins2025} for further details). The LRD stack from \citet{Delvecchio2025}, which we multiplied by 1.4 to match the optical and near-infrared emission of the \citet{Perez26} stack, is similar in shape to the other two LRD stacks, also showing a rise in the near-infrared continuum. Therefore, our reddened accretion disc + torus model also accommodates them. %We emphasize that the torus models shown in Fig. \ref{fig4} were not fitted to the LRD stacks, but to the $\sim$3-27 $\mu$m Sy1 stack. 

In Fig. \ref{fig4} we show the SED of the local LRD LORD/The Egg \citep{Ji2026,Lin2026}. Scaling its W4 flux (22 $\mu$m) to the Sy1 stack,  the W1 (3.4 $\mu$m) and W2 (4.6 $\mu$m) fluxes fall within its standard deviation, while the W3 flux (12 $\mu$m) is above it, possibly due to a contribution from emission features (see Section \ref{lord}). Although in F$_{\lambda}$ units the SED of LORD does not appear different from the \citet{Akins2025} LRD stack (see Fig. 1 in \citealt{Ji2026}), from Fig. \ref{fig4} it becomes clear that the V-shape is more pronounced and the fluxes between $\sim$0.7 and 1 $\mu$m are above the high-z LRD stacks, partly due to contamination from emission lines. In any case, it is normal for individual sources to differ from stacks of hundreds of galaxies, as the stacking process washes out source peculiarities. We refer the reader to \citet{Madau2026Lord} for an example of an alternative accretion disc model and a clumpy dusty screen successfully replicating the LORD SED.

\subsection{Energy balance arguments}
\label{enbal}

\citet{Lin2026} used three BBs of T$\sim$1000, 300, and 90 K to reproduce the infrared SED of LORD/Egg, and a fourth BB of T$\sim$5000 K to account for the optical continuum. As can be seen from their Table D1, the integrated luminosity of the three infrared BB components is almost twice the luminosity of the 5000 K BB. Therefore, in the scenario in which the optical emission is intrinsically red and not dust-reddened/extincted (i.e. the BH-star/cocoon scenarios), then the dust covering factor or RE ($\rm L_{dust}/L_{bol} \sim L_{dust}/L_{opt}$ in the BH-star case) would be larger than unity. This model-independent infrared versus optical energy balance issue represents a challenge for the previously mentioned scenarios.

In the case of the accretion disc+torus model shown in Fig. \ref{fig4} (reddened model in Table \ref{tab:lum}), the energy of the dust component is almost three times higher than that of the apparent optical component (disc), but the key difference with the BH-star model is that in our case the optical corresponds to a reddened (by $\rm A_V\simeq2$ mag) accretion disc component -- the intrinsic bolometric luminosity is significantly higher. This means that if we consider the unreddened model (see Table \ref{tab:lum}), the luminosity of the dust component is lower than that of the disc ($\rm L_{dust}/L_{disc}$=0.17), resulting in a reasonable dust covering factor. %, not too dissimilar from standard AGN. 

Our proposed scenario of reddened accretion disc and low RE or covering factor torus to explain the optical-to-infrared SED of LRDs is supported by the recent work of \citet{Chandna2026}. They used the SKIRT radiative transfer code to post-process zoom-in cosmological simulations and compared their models to LRD observations, including the stack of \citet{Perez26}. They reported that the steep red optical slope of LRDs arises from a dust-attenuated AGN, whereas a less obscured stellar component dominates the UV emission. A similar approach, which includes a more detailed geometrical radiative transfer model, was proposed by \cite{Madau2026b} and, for the specific case of the Egg/LORD, by \cite{Madau2026Lord}, reaching similar results as our more phenomenological approach.

\begin{table}
\caption{Integrated luminosities of the accretion disc and torus components of the model SED in Fig. \ref{fig4}, with and without extinction.}
\label{tab:lum}
\centering
\begin{tabular}{lcccc}
\hline\hline
Model $\rm C_T=0.08$ & $L_{\rm disc}$ & $L_{\rm dust}$ & $L_{\rm total}$ & $L_{\rm dust}/L_{\rm disc}$ \\
 & \multicolumn{3}{c}{($10^{44}$\,erg\,s$^{-1}$)} & \\
\hline
Reddened ($A_V=1.9$\,mag) & 0.87 & 2.47 & 3.34 & 2.8 \\
Unreddened ($A_V=0$)      & 17.4 & 2.95 & 20.4 & 0.17 \\
\hline
\end{tabular}
\tablefoot{Luminosities are integrated over $0.01$--$36000\,\mu$m ($z=7$, $d_L=69$\,Gpc), using the SMC bar extinction law for the reddened model. The proxy of \citet{Maiolino07}, i.e. Eq. \ref{eq1}, gives $C_T=0.08$ for the unreddened model, whereas directly integrating the disc and torus components gives 0.17. This mismatch comes from the conversion factors used in Eq. \ref{eq1}.}
\end{table}

\subsection{Dust masses}
\label{dust}

If the torus model shown in Fig. \ref{fig4} (black dot-dashed line) indeed resembles the rest-frame mid- and far-infrared emission of high-redshift LRDs, the 5$\sigma$ upper limits from \citet{Akins2025} would be largely consistent with our analysis. Their Herschel/PACS upper limits at 70 and 160 $\mu m$ ($\sim$10 and 23 $\mu m$ rest-frame) are $\sim$90 and 60 times above the flux of the scaled Sy1 stack and torus model, respectively. The upper limit at $\rm \lambda_{rest} > 100~\mu m$ from \citet{Delvecchio2025} is more restrictive than those from \citet{Akins2025}, but still well above the torus emission at those wavelengths (see Fig. \ref{fig4}). % , which strongly declines at $\rm \lambda_{rest}$>20-30 $\mu m$. 
This comparison shows that the rest-frame infrared non-detections obtained for high-redshift LRDs should not be interpreted as a lack of dust in these targets. The increasing temperature of the CMB reduces the contrast between dust and the CMB emission \citep{daCunha2013}, making it difficult to obtain constraining upper limits even at ALMA sensitivities \citep{Casey2024,Casey25}.

We fitted the dust torus model in Fig. \ref{fig4} (dot-dashed black line) with a sum of four modified BBs ($L_\nu=4\pi M_{\rm dust}\kappa_\nu B_\nu(T)$, with $\kappa_\nu=\kappa_{100}\,(100\,\mu{\rm m}/\lambda)^{\beta}$; see Fig.~\ref{fig:BBs}) of T$\simeq$100, 220, 570, and 1175 K. We adopted $\beta=1.7$, set by the slope of the long-wavelength tail ($\lambda F_\lambda\propto\lambda^{-4.65}$), and $\kappa_{100}=25$\,cm$^2$\,g$^{-1}$ \citep{Draine2003}. The fit reproduces the torus  SED at $\lambda\geq1\,\mu$m with an rms of 0.05\,dex. More than 99\% of the dust mass is in the coolest component ($T\simeq100$\,K), giving $M_{\rm dust}\simeq1.4\times10^4\,M_\odot$. Varying $\beta$ between 1.5 and 2 and $\kappa_{100}$ between 10 and 60\,cm$^2$\,g$^{-1}$ changes this value between $\sim4\times10^3$ and $\sim6\times10^4\,M_\odot$. Since the cold component is constrained only by the far-infrared tail of the torus model, which is an extrapolation beyond the data ($\lambda\lesssim4\,\mu$m), we regard this value as an order-of-magnitude estimate. Regardless, our dust mass estimation is in excellent agreement with the average dust mass predicted by \citet{Casey2024}, of $M_{\rm dust}=1.6\pm^{4.8}_{0.9}\times10^4\,M_\odot$, for 675 LRDs at z$\gtrsim$4. They used the available measurements of $\rm A_V$, $\rm M_{UV}$, and galaxy effective radius ($\rm R_{eff}$) to derive dust mass and temperature, the latter being $\rm T_{dust}\simeq$110 K. Most importantly, the dust masses inferred by our modelling are consistent with the range of dust masses typical of AGN dusty tori \citep[$10^2-10^4~M_\odot$][]{GarciaBurillo21,Hatziminaoglou2008}.

The high-redshift LRD stack from \citet{Delvecchio2025}, shown as green dots in Fig. \ref{fig4}, was reproduced by \citet{Madau2026b} using a dust covering factor of 0.2 and a reddened super-Eddington AGN continuum with $\rm A_V\simeq3$ mag. They reported a dust mass of $\sim$60-120 solar masses for that torus, assuming a gas-to-dust ratio of 1000, which is probably more appropriate for metal-poor hosts such as those of LRDs \citep{Madau2026b}. The dust mass that we derived from the torus component in Fig. \ref{fig4}, of $M_{\rm dust}\simeq10^4\,M_\odot$, is 
two orders of magnitude above the $\sim10^2\,M_\odot$ of \citet{Madau2026b}. The difference is most likely due to the different modelling of the cold component of the dust emission, which is poorly constrained due to the lack of tight constraints at longer wavelengths, especially taking into account the steep dependence of the dust emission on the temperature. The torus modelling is also significantly different between the two approaches (especially in terms of density profile). Our estimated dust mass is two orders of magnitude below the current upper limits for high-redshift LRDs ($M_{\rm dust}\lesssim10^6\,M_\odot$; \citealt{Casey25,Chen2025,Setton2025}). %Our dust mass estimate agrees with the prediction from $\rm A_V$ and $\rm R_{eff}$ of LRDs at z$\gtrsim$4 \citep{Casey2024}.

\section{Conclusions}

We used rest-frame 3--27\,$\mu$m JWST spectra of eight local Sy1 galaxies to make predictions on the dust emission of LRDs. We normalised the individual Sy1 spectra by their broad H$\alpha$ and 5100\,\AA\ continuum luminosities and stacked them, obtaining $\lambda L_\lambda/L_{\rm H\alpha}\simeq30$--60 and $\lambda L_\lambda/\lambda L_{5100}\simeq0.8$--1.8 across 3--27\,$\mu$m. 

We tested these scalings with the NIRSpec+MIRI stack of 249 LRDs at $2.3<z<9.3$ from \citet{Perez26} and with the SED of a local LRD at $z=0.1$ \citep{Ji2026,Lin25}. The 5100\,\AA\ continuum gives the more reliable prediction on the infrared emission of LRDs (rest-frame $\lambda\lesssim4$\,$\mu$m for the high-z stack and $\lambda\lesssim22$\,$\mu$m for the local LRD) when modest extinction is considered ($A_V\lesssim1$\,mag). Although the high-z stack and, to a lesser extent, the local LRD SED probe large areas that may include host-galaxy emission, this agreement is consistent with an AGN-dominated optical continuum.

%In Section \ref{LRDs_normalizations} we showed that the infrared emission of LRDs can be predicted from their optical continuum luminosity using the Sy1 stacks shown in Fig. \ref{fig3}. This approach has the advantage of being model-independent. However, the torus+accretion disc model that fits both the Sy1 stack and optical point (see Fig. \ref{fig2}) does not reproduce the spectral shape of LRDs, suggesting that their torus properties differ from those of Sy1s. We therefore attempt to reproduce the optical-to-infrared spectral shape of high-redshift LRDs by varying the torus covering factor (i.e., the scaling of the accretion disc relative to the torus model) and foreground extinction.

%Although, based on our results, it is possible to predict the infrared emission of LRDs from their 5100\,\AA\ continuum emission, the  torus+accretion disc model that fits the local Sy1 stack does not simultaneously reproduce the optical and infrared continuum shape of LRDs. This requires scaling up the accretion disc contribution relative to the torus, which results in a lower covering factor ($C_{\rm T}\simeq0.1$) than inferred for local Sy1 galaxies, and then redden it by $A_V\simeq2$\,mag using the SMC bar extinction law. This combination offers an alternative to BH-star scenarios, which explain the optical continuum as an unreddened blackbody of $T\sim5000$\,K, and yields a more plausible infrared-to-optical energy balance. %, suggesting that LRDs are not dust deficient.

Although our results indicate that the infrared emission of LRDs can be predicted, within a factor of $\sim$1.5-2, from their 5100\,\AA\ continuum, the torus+accretion disc model that fits the local Sy1 stack does not simultaneously reproduce the optical and infrared SED of LRDs. Reproducing both requires a more reddened accretion disc ($A_V\simeq2$\,mag) and a torus with a lower covering factor or RE ($C_{\rm T}\simeq0.1-0.2$) than inferred for local Sy1 galaxies. 
%scaling up the accretion disc relative to the torus, which corresponds to a lower covering factor ($C_{\rm T}\simeq0.1$) than inferred for local Sy1 galaxies, and reddening it by $A_V\simeq2$\,mag with the SMC bar extinction law. 
This combination offers an alternative to BH-star scenarios (which ascribed the optical continuum to an unreddened blackbody of $T\sim5000$\,K), while yielding a more plausible infrared-to-optical energy balance.

From the fit of the torus model that reproduces the SED of the high-redshift LRD stack of \citet{Perez26} with a sum of four optically thin modified BBs (T$\simeq$100, 220, 570, and 1175 K), we derive a dust mass of $\sim10^4\,M_\odot$ (between $\sim4\times10^3$ and $\sim6\times10^4\,M_\odot$ considering different values of $\beta$ and $\kappa_{100}$). This dust mass is in good agreement with the prediction from \citet{Casey2024}, of $M_{\rm dust}=1.6\pm^{4.8}_{0.9}\times10^4\,M_\odot$, obtained from measured values of $\rm A_V$, $\rm M_{UV}$, and $\rm R_{eff}$ for 675 LRDs at z$\gtrsim$4. This dust masses are two orders of magnitude below the current upper limits for high-redshift LRDs, based on their ALMA non detections ($M_{\rm dust}\lesssim10^6\,M_\odot$; \citealt{Casey25,Chen2025,Setton2025}). According to the results presented here, LRDs are not dust deficient, but their accretion disc and torus properties would be different from those of local AGN.

\begin{acknowledgements}
This work is based on observations made with the NASA/ESA/CSA James Webb Space Telescope. The data were obtained from the Mikulski Archive for Space Telescopes at the Space Telescope Science Institute, which is operated by the Association of Universities for Research in Astronomy, Inc., under NASA contract NAS 5-03127 for JWST and from the European JWST archive (eJWST) operated by the ESAC Science Data Centre (ESDC) of the European Space Agency. These observations are associated with programs GO 2016, 3535, 4972, 7195, and the Early Release Science Program 1328. CRA thanks Thaisa Storchi-Bergmann and Allan Schnorr-M{\"u}ller for their help with the BLR extinction values. 
CRA and AA acknowledge support from the Agencia Estatal de Investigaci\'on of the Ministerio de Ciencia, Innovaci\'on y Universidades (MCIU/AEI) under grants ``Tracking active galactic nuclei feedback from parsec to kiloparsec scales'', with reference PID2022$-$141105NB$-$I00, and ``Local windows into the most extreme phases of black hole growth'', with reference PID2025$-$176305NB$-$I00, and the European Regional Development Fund (ERDF). This work is part of grant CEX2025-001609-S, awarded to the IAC under the Severo Ochoa Centre of Excellence program and funded by MICIU/AEI/10.13039/501100011033. AAH and LHM acknowledge support from grant PID2021-124665NB-I00 funded by MCIN/AEI/10.13039/501100011033 and by ERDF A way of making Europe. NAL acknowledges support from the Fundación Occident and the Instituto de Astrofísica de Canarias under the Visiting Researcher Programme 2025-2028 agreed between both institutions. PGP-G acknowledges support from grants PID2022-139567NB-I00 and PID2025-169195NB-I00 funded by Spanish Ministerio de Ciencia, Innovaci\'on y Universidades MICIU/AEI/10.13039/501100011033, and the European Union FEDER program {\it Una manera de hacer Europa}. AA is funded by the European Union (Widening Participation, ExGal-Twin, GA 101158446). EB acknowledges support from the Spanish grant PID2022-138621NB-I00 and PID2021-123417OB-I00, funded by MCIN/AEI/10.13039/501100011033/FEDER, EU. MS acknowledges support by the Ministry of Science, Technological Development and Innovation of the Republic of Serbia (MSTDIRS) through contract no. 451-03-33/2026-03/200002 with the Astronomical Observatory (Belgrade).

\end{acknowledgements}

\bibliographystyle{aa} 
\bibliography{GATOS_Sy1} % if your bibtex file is called example.bib

\begin{appendix} 
\section{Comparison between JWST NIRSpec+MIRI and Spitzer/IRS spectra of the targets}\label{appendix}

\begin{figure*}[!h]
  \centering
{\par\includegraphics[width=0.85\columnwidth]{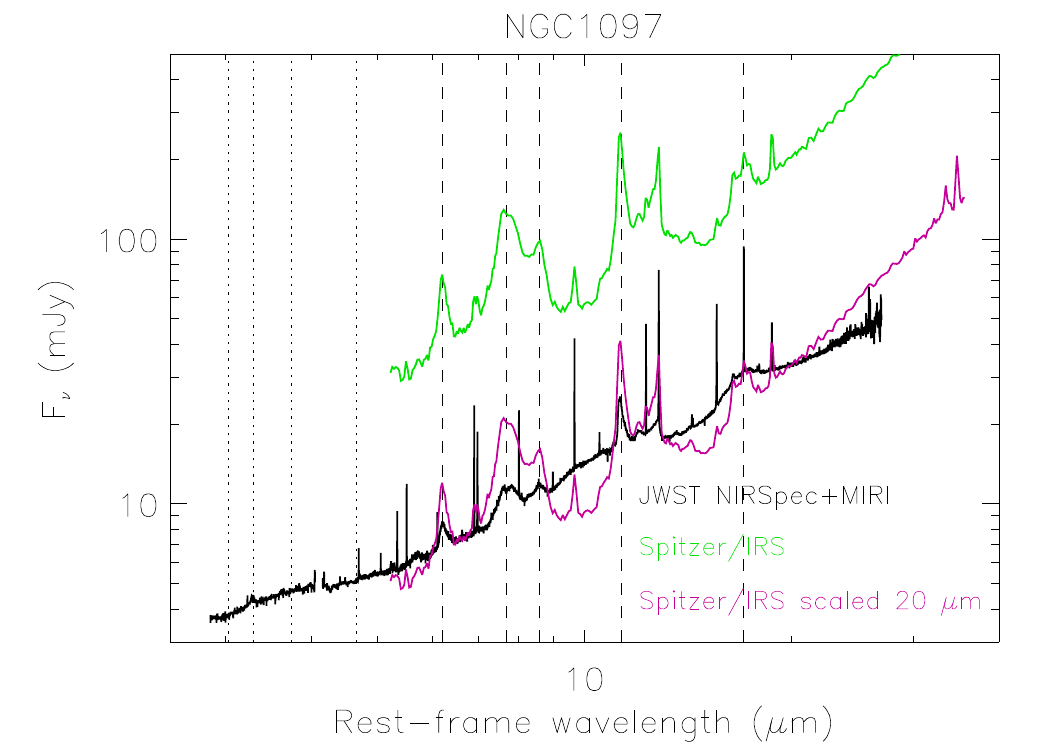}
\includegraphics[width=0.85\columnwidth]{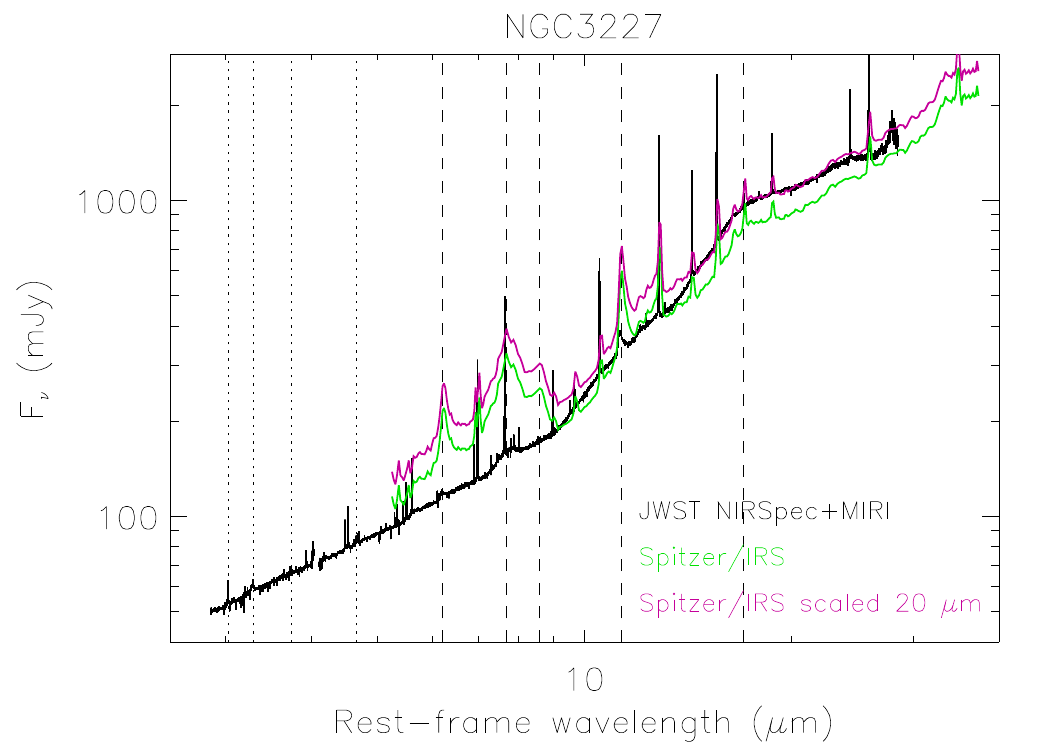}
\includegraphics[width=0.85\columnwidth]{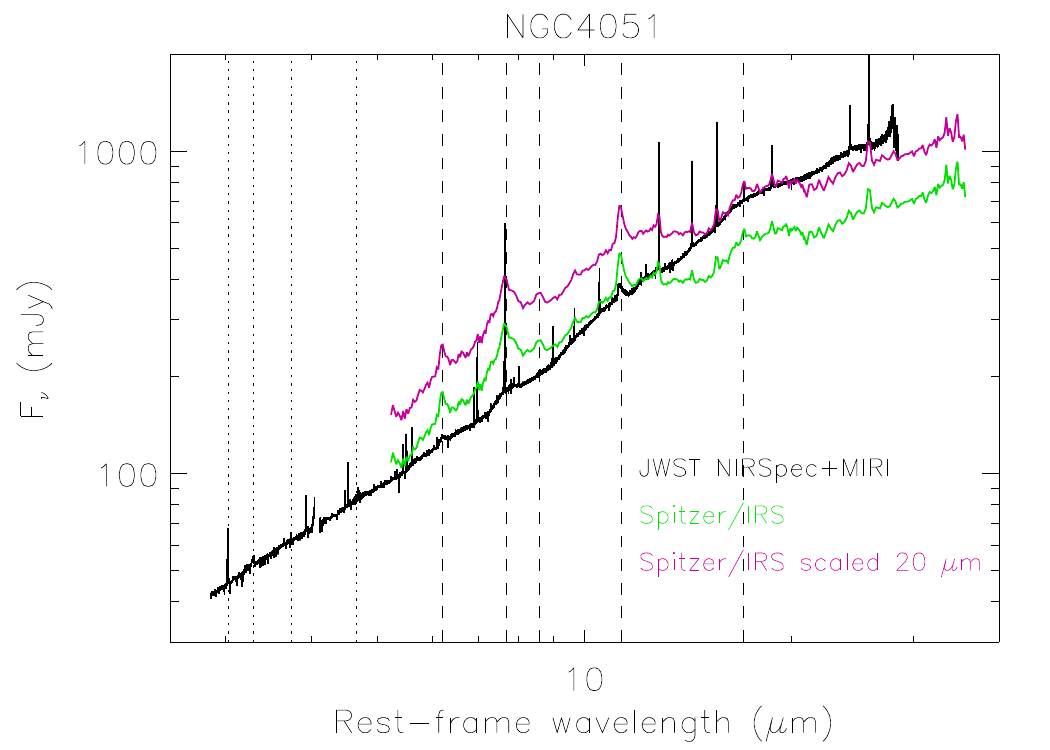}
\includegraphics[width=0.85\columnwidth]{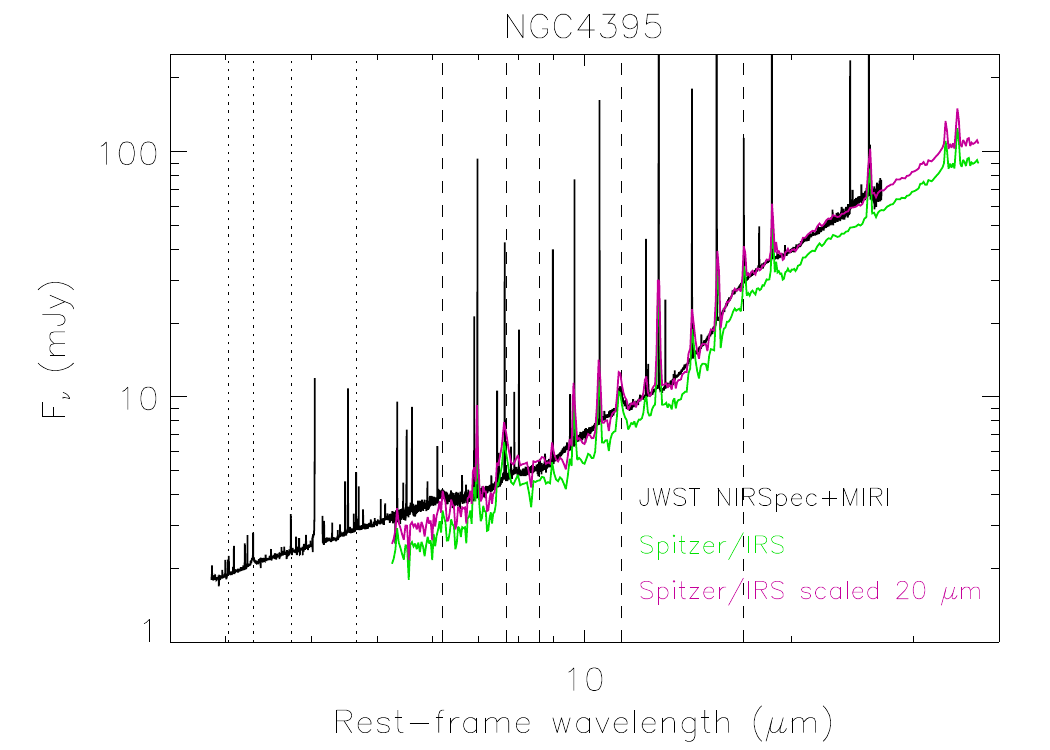}
\includegraphics[width=0.85\columnwidth]{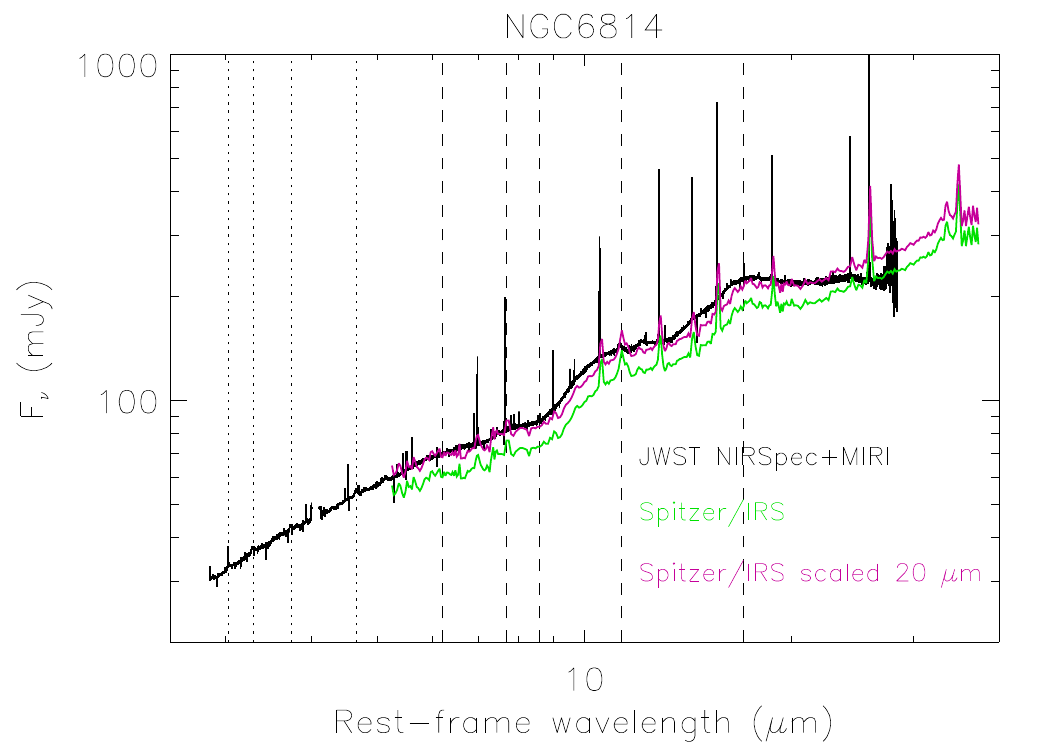}
\includegraphics[width=0.85\columnwidth]{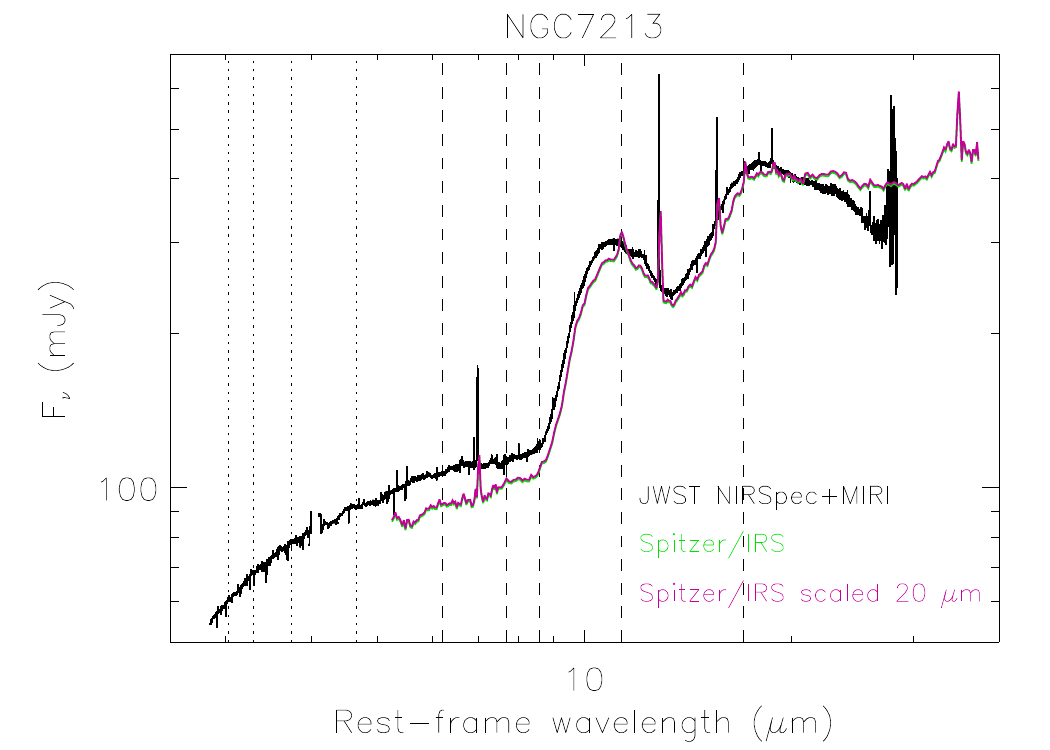}
\includegraphics[width=0.85\columnwidth]{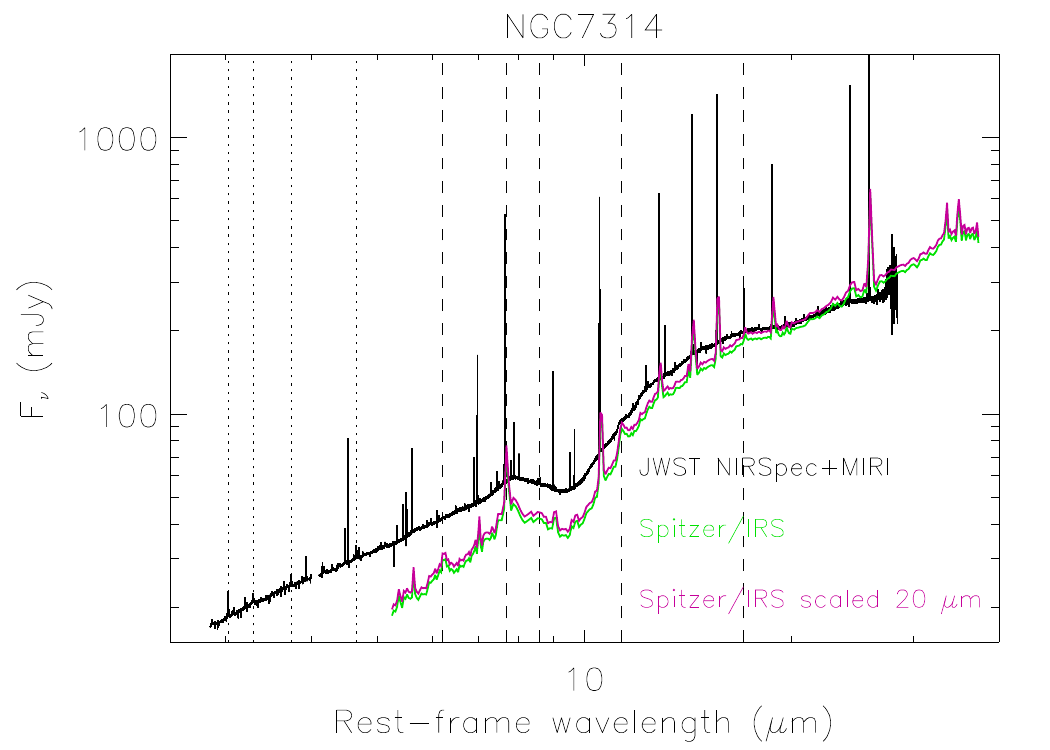}
\includegraphics[width=0.85\columnwidth]{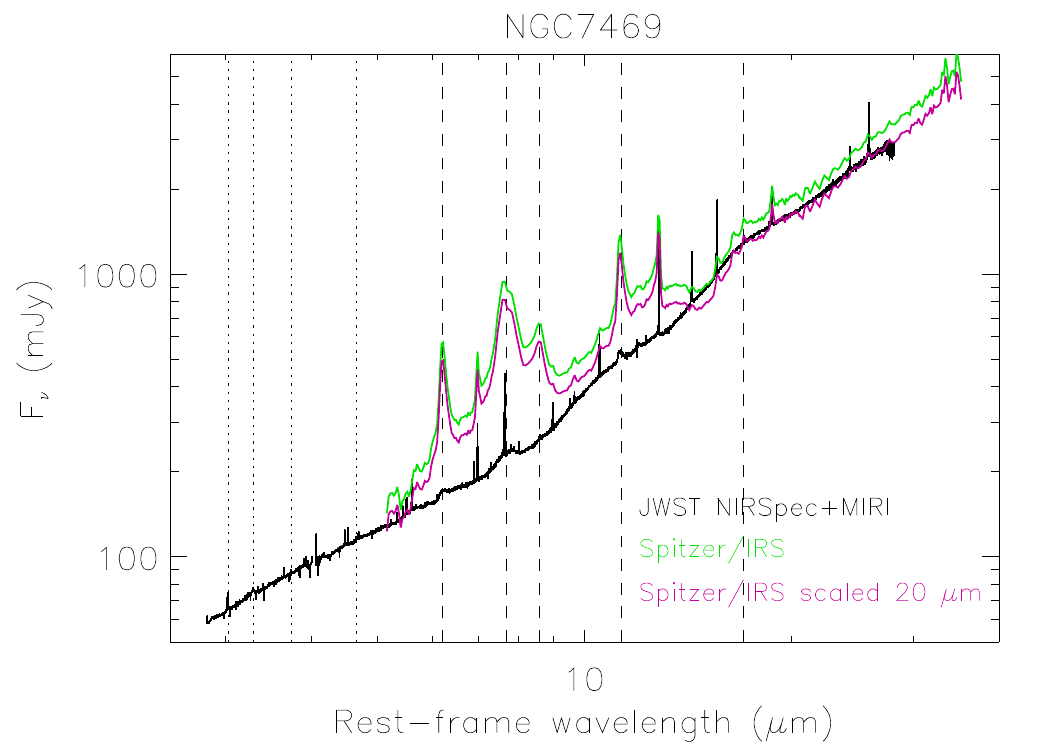}\par}
\caption{Rest-frame JWST/NIRspec+MIRI spectra of the eight Seyfert 1 galaxies (black line). The CASSIS optimal extraction Spitzer/IRS spectra is shown in green, and the purple line corresponds to the same spectra but scaled to the JWST/NIRspec+MIRI (at 20 $\mu$m). Vertical dotted lines indicate the wavelength of the Pfund $\beta$, $\gamma$, $\delta$, and $\epsilon$ lines. Vertical dashed lines indicate the location of the 6.2, 7.7, 6.8, 11.3, and 17 $\mu$m PAH bands.}
              \label{figA1}
    \end{figure*}

Figure \ref{figA1} shows a comparison between the nuclear JWST NIRSpec+MIRI used in this work and their Spitzer/IRS spectra. The IRS spectra was dowloaded from the Combined Atlas of Sources with \textit{Spitzer} IRS Spectra (\href{https://cassis.sirtf.com/atlas/welcome.shtml}{CASSIS}). The observations were done using the Short-Low (SL) and Long-Low (LL) spectrograph modules, which cover the wavelength range 5.3--38 $\mu$m with a spectral resolution of R = $\lambda$/$\Delta\lambda\sim$60--130. The SL1 and SL2 slits have widths of 3.6 and 3.7\arcsec, and the LL slits of 10.5 and 10.7\arcsec. Since the targets are type-1 AGN, optimal extraction was chosen.

\section{Torus models}\label{appendixC}

\begin{figure*}[!h]
  \centering
%{\par\includegraphics[width=1.0\columnwidth]{hoenig_interpol_stack2_stdp_Av0_residuals.pdf}
%\includegraphics[width=1.0\columnwidth]{nenkova_interpol_stack2_stdp_Av0_residuals.pdf}\par}
\includegraphics[width=2.0\columnwidth]{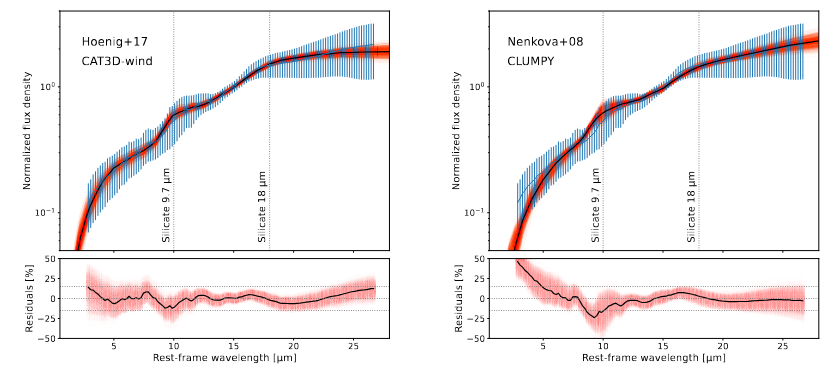}
  \caption{Smoothed and interpolated mean Sy1 stack (solid blue line) and fits the with the CAT3D-wind and CLUMPY torus models (shaded orange area: model SEDs between the 16 and 84 percentiles of the posterior. Solid black line: 50 percentile of the posterior). Vertical dotted lines indicate the peak wavelengths of the silicate features. The bottom panel shows the residuals as a percentage of the continuum flux, with the horizontal dotted lines indicating 0 and 15\% error for guidance.}%No AGN, Av=0. The names of the files are the ones commented above figC1.png}
              \label{figC1}
    \end{figure*}

Figure \ref{figC1} shows the Sy1 stack (see right panel of Fig. \ref{fig1}), in blue, fitted with the CAT3D-wind and CLUMPY torus models (orange and black lines). 

\section{Normalizations of individual galaxies and average normalized fluxes in JWST filters}\label{appendixB}

\begin{figure*}[!h]
  \centering
 {\par\includegraphics[width=1.0\columnwidth]{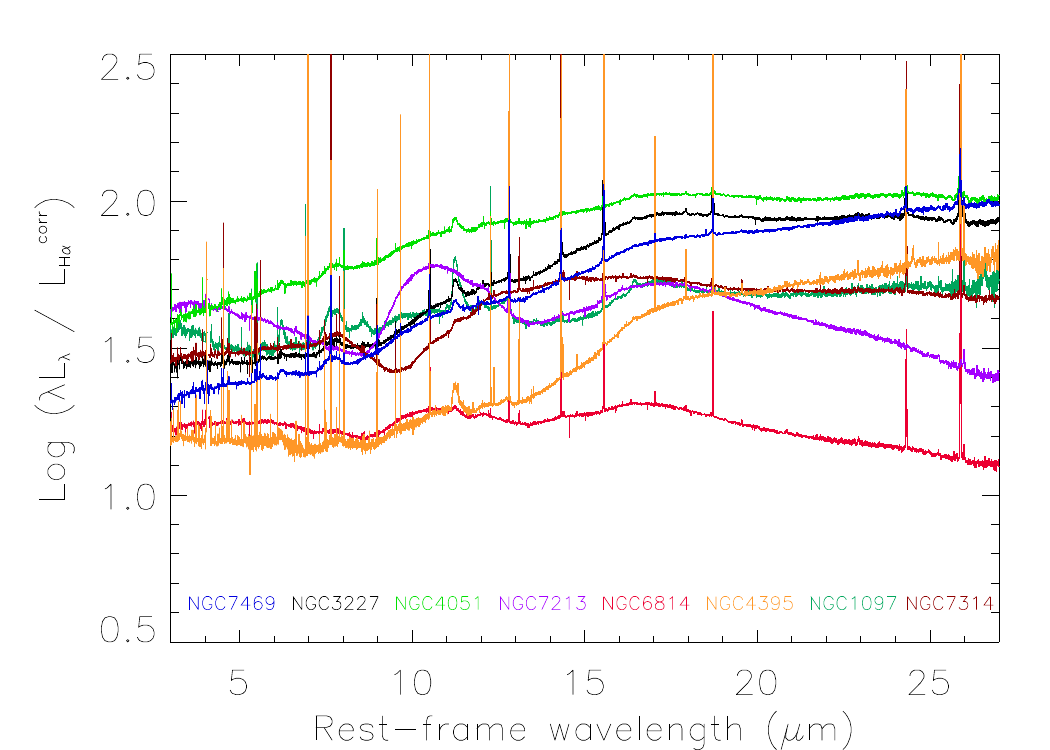}
  \includegraphics[width=1.0\columnwidth]{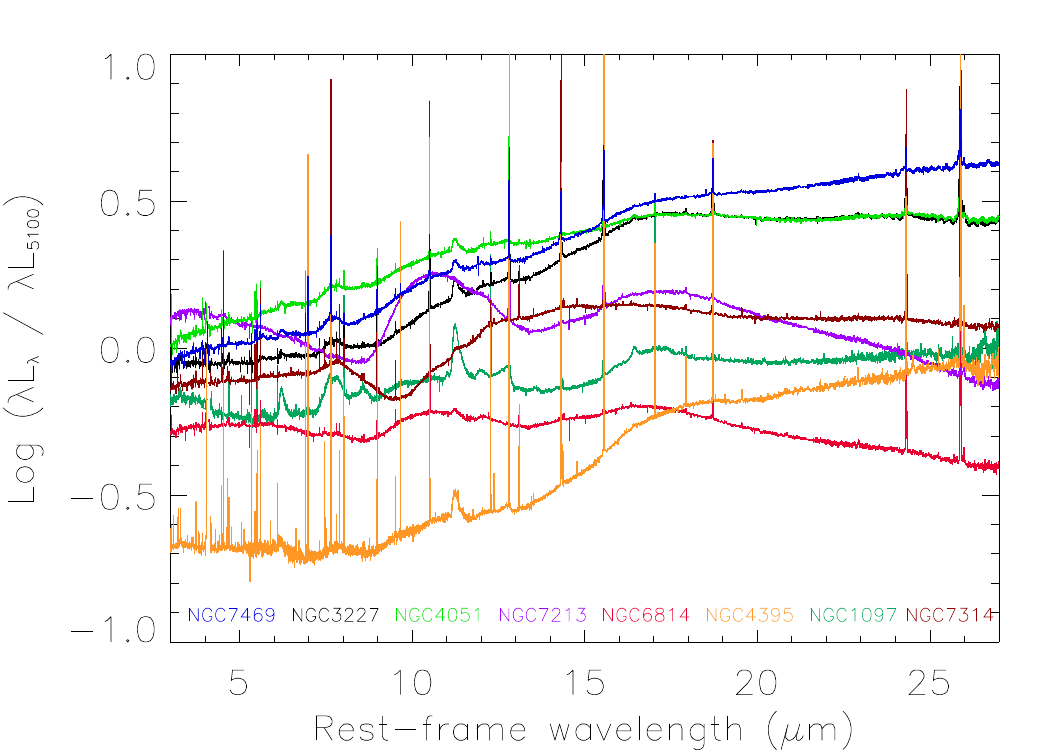}\par}
  \caption{Rest-frame infrared spectra of the Sy1 galaxies, with same colors as in the left panel of Fig. \ref{fig1}, normalized to their corresponding H$\alpha$ (left panel) and 5100 \AA~luminosities (right panel).}
              \label{figB1}
    \end{figure*}

Figure \ref{figB1} shows the rest-frame infrared spectra of the Sy1 galaxies, normalized to their corresponding H$\alpha$ (left panel) and 5100 \AA~luminosities (right panel). The 5100 \AA~luminosities were calculated from $\rm L_{H\alpha}^{avg}$ in Table \ref{tab2} and Eq. 1 in \citet{Greene2005}. The H$\alpha$ luminosities of NGC\,6814 and NGC\,7314 were extinction-corrected using the values reported in Table \ref{tab2}, and then used to calculated the corresponding L$_{\rm 5100}$.

\begin{table}[h!]
\centering
\footnotesize
  \caption{Average normalized fluxes in NIRCAM and MIRI wide filters.} 
\begin{tabular}{lccccc}
\hline
\hline
Filter & $\lambda$ & $\lambda_b$ & $\lambda_r$ & $\rm log (\frac{\lambda L_{\lambda}}{L_{H\alpha}})$ &  $\rm log (\frac{\lambda L_{\lambda}}{L_{5100}})$ \\
  & ($\mu$m) & ($\mu$m) & ($\mu$m) & & \\
  \hline
 F356W & 3.565 &  3.136  &  3.981   &     1.46842  &   -0.09200   \\ 
 F444W & 4.402 &  3.881  &  4.982   &     1.46868  &   -0.08839   \\	
 F560W & 5.6   &  5.054  &  6.171   &     1.47254  &   -0.08317   \\ 
 F770W & 7.7   &  6.581  &  8.687   &     1.49753  &   -0.05972   \\ 
F1000W & 10.0  &  9.0230 &  10.891  &     1.58433  &    0.03371   \\ 
F1130W & 11.3  &  10.953 &  11.667  &     1.65242  & 	0.10138   \\ 
F1280W & 12.8  &  11.588 &  14.115  &     1.65717  & 	0.10856   \\ 
F1500W & 15.0  &  13.527 &  16.640  &     1.72438  & 	0.17832   \\ 
F1800W & 18.0  &  16.519 &  19.502  &     1.78963  & 	0.24330   \\ 
F2100W & 21.0  &  18.477 &  23.159  &     1.77976  & 	0.23429   \\ 
F2550W & 25.5  &  23.301 &  26.733  &     1.80168  & 	0.25419   \\ 
\hline 
\end{tabular}
\tablefoot{Columns 1-4 are the name, pivot , blue, and red wavelengths of the JWST NIRCAM and MIRI wide filters. We refer the reader to the instruments webpages for details on their definition. Columns 5 and 6 are the average fluxes measured between the blue and red wavelengths of each filter in the normalized stack spectra shown in Fig. \ref{fig3}.}
\label{tab4} 
\end{table}

\section{Bolometric corrections}
\label{bolometric}

\begin{figure*}[!h]
  \centering
  {\par\includegraphics[width=1.0\columnwidth]{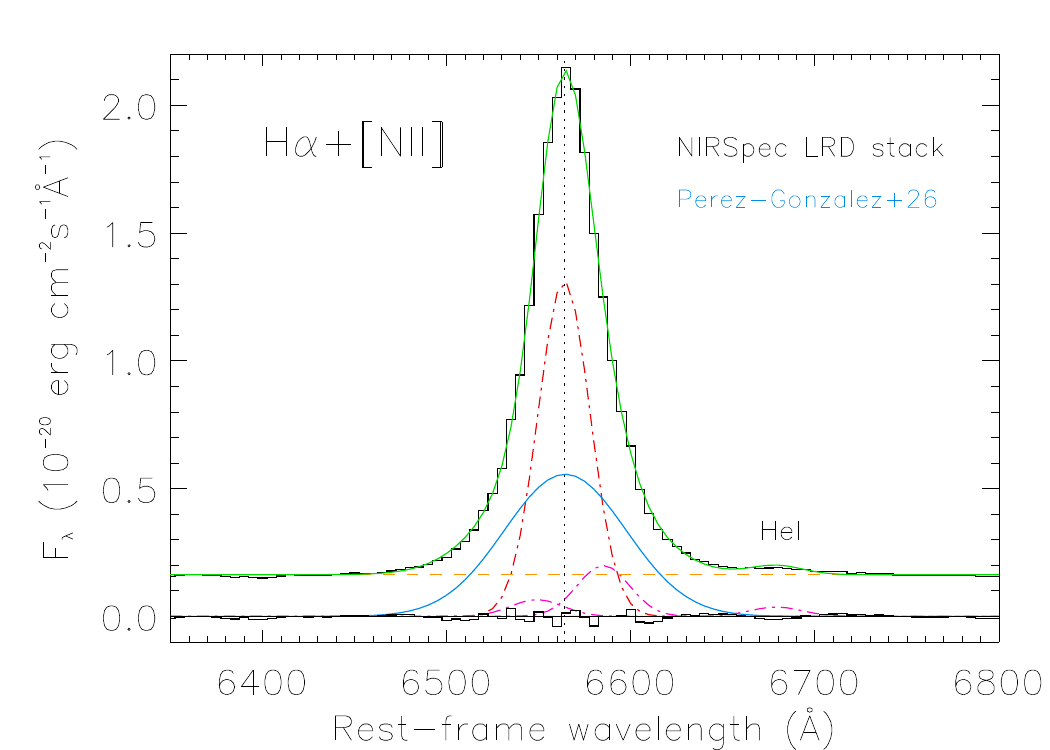}
 \includegraphics[width=1.0\columnwidth]{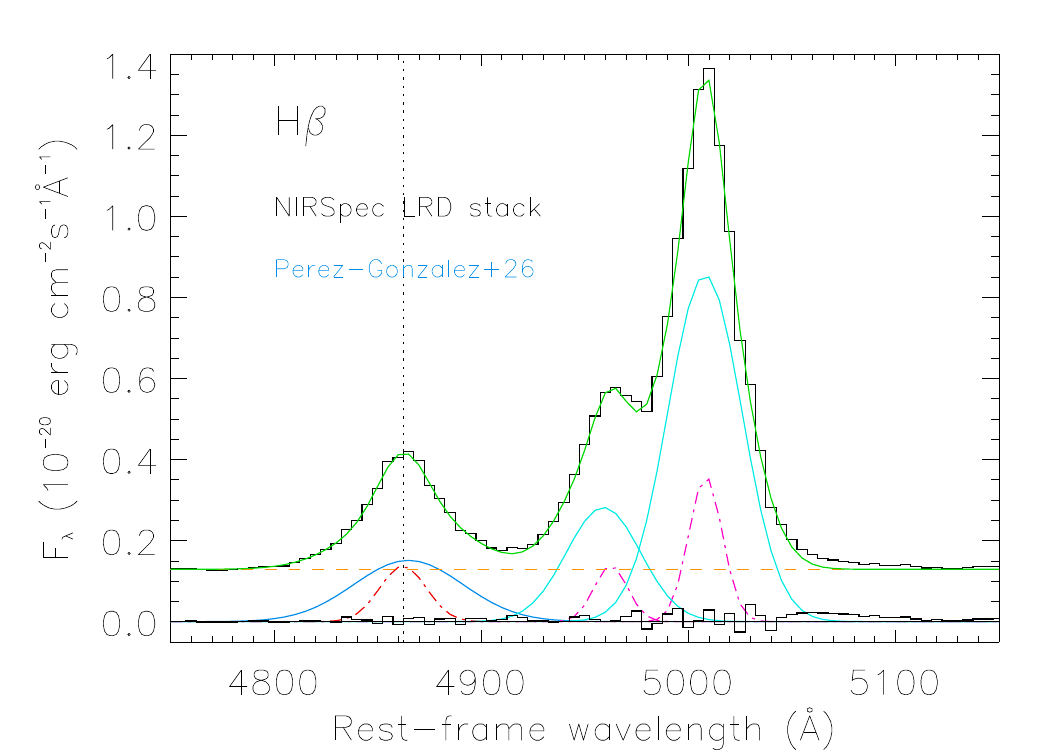}\par}
  \caption{Rest-frame NIRSpec LRD stack from \citet{Perez26} around the H$\alpha$ (left panel) and H$\beta$ and [OIII] emission lines (right panel). The stacked H$\alpha$ and H$\beta$ profiles were reproduced with two Gaussian components (red and blue lines), both of them resolved at the spectral resolution of the NIRSpec/PRISM in the case of H$\alpha$ (see Table \ref{tab5}). % ($\sim$1100 km~s$^{-1}$).
  The [NII] and HeI$\lambda$6678 \AA~lines are shown in pink, and their FWHMs have been fixed to that of the intermediate component of H$\alpha$ ($\sim$1050 km~s$^{-1}$). The [OIII] lines were fitted with two components, shown in pink and cyan. In both panels, the continuum is shown as a dashed orange line, the fit as a green solid line, and the residuals in black.}
              \label{figD2}
    \end{figure*}

We measured the broad H$\alpha$ flux from the fit shown in the left panel of Fig. \ref{figD2}. We used two Gaussian components, of FWHM=3400 (broad) and 1050 km~s$^{-1}$ (intermediate), typical of high-redshift LRDs  \citep{Matthee24,Greene2024,Brazzini26}, both resolved at the PRISM resolution (see Table \ref{tab5}). More complex profiles have been tested with higher spectral resolution data (see \citealt{Geris2026} and references therein), but at the PRISM resolution we rather keep the fit as simple as possible. We fitted two additional Gaussians to account for the [NII] emission, having the same FWHM as the ``narrow'' (intermediate, hereafter) H$\alpha$ component (see left panel of Fig. \ref{figD2}). HeI$\lambda$6678 \AA~is also detected in the LRD stack, overlapping with the red wing of H$\alpha$. The fluxes and luminosities of the broad and intermediate H$\alpha$ components are reported in Table \ref{tab5}.

Using the integrated H$\alpha$ fluxes and the fitted continuum (orange dashed line in Fig. \ref{figD2} and Table \ref{tab5}), we measure EWs of 282 and 285 \AA~for the broad and intermediate components, which are within the range of nearby AGN (EW=[15,500] \AA~for $\rm log~L_{H\alpha}^b$=[40, 44] erg~s$^{-1}$; \citealt{Stern2012}). If we sum the two H$\alpha$ components (red and blue lines in the left panel of Fig. \ref{figD2}) we measure an EW=567 \AA, and if we also include the [NII] line fluxes, EW=625 \AA. The values considering the whole H$\alpha$ and H$\alpha$+[NII] profile are typical of high-redshift LRDs ($\sim$400-1000 \AA; \citealt{Lin2024,Setton2025,Brazzini26,Geris2026b}). Instead, the EW of the broad H$\alpha$ component alone would imply that the covering factor of the BLR might not be that different from that of local AGN (see also \citealt{Geris2026}). We refer the reader to theoretical and observational studies of type-1 AGN at z$\gtrsim$3 using high spectral resolution data (e.g. \citealt{Brazzini2025,Ignas2026,Geris2026,Matthee2026,Madau2026,Scholtz2026}) for further discussion on the evolution of the structure and properties of the BLR with cosmic time.

    \begin{table}[h!]
\centering
\footnotesize
  \caption{Optical spectral properties of NIRSpec LRD stack.} 
\begin{tabular}{lcc}
\hline
\hline      
 & Broad & Interm. \\
 \hline
 $\rm \lambda_{H\alpha}$ ($\AA$)  & 6564.5$\pm$0.4 & 6563.8$\pm$0.2 \\
 $\rm FWHM_{H\alpha}$ (km~s$^{-1}$)   & 3380$\pm$110  & 1050$\pm$30 \\
 $\rm EW_{H\alpha}$ ($\AA$)  & 282 & 285  \\
 $\rm F_{H\alpha}$ ($\rm erg~cm^{-2}~s^{-1}$) & (3.68$\pm$0.19)$\times10^{-18}$ & (3.72$\pm$0.17)$\times10^{-18}$ \\
 $\rm L_{H\alpha}$ ($\rm erg~s^{-1}$)         & (2.10$\pm$0.11)$\times10^{42}$ & (2.12$\pm$0.09)$\times10^{42}$ \\
 \hline                      
 $\rm \lambda_{H\beta}$ ($\AA$)  & 4864.9$\pm$1.5 & 4862.3$\pm$0.9  \\
 $\rm FWHM_{H\beta}$ (km~s$^{-1}$)   & 3400$\pm$430 & \dots  \\
  $\rm EW_{H\beta}$ ($\AA$)  & 77 & 30 \\
 $\rm F_{H\beta}$ ($\rm erg~cm^{-2}~s^{-1}$) & (7.98$\pm$1.17)$\times10^{-19}$ & (3.10$\pm$1.27)$\times10^{-19}$\\
 $\rm L_{H\beta}$ ($\rm erg~s^{-1}$)         & (4.55$\pm$0.67)$\times10^{41}$ & (1.77$\pm$0.72)$\times10^{41}$ \\
 \hline
$\rm \lambda F_{5100}$ ($\rm erg~cm^{-2}~s^{-1}$) &
\multicolumn{2}{c}{5.69$\times10^{-17}$} \\
$\rm \lambda L_{5100}$ ($\rm erg~s^{-1}$) &
\multicolumn{2}{c}{3.23$\times10^{43}$} \\
\hline
\end{tabular}
\tablefoot{Central wavelength, FWHM corrected from instrumental broadening ($\sim$1100 km~s$^{-1}$ for H$\alpha$ and $\sim$1750 km~s$^{-1}$ for H$\beta$, using the median resolution values from \citealt{Perez26}, corrected by a factor of 1/0.70), EW, integrated flux, and luminosity of the H$\alpha$ and H$\beta$ broad and intermediate components (blue and red lines in Fig. \ref{figD2}). The line fluxes were multiplied by a factor of (1+z), with z=7, because they were measured in the rest-frame spectrum, in flux density units (F$_{\lambda}$). The 5100 \AA~flux was measured as the average of the $\lambda F_{\lambda}$ rest-frame spectrum between 5080 and 5120 \AA, so no (1+z) factor was applied.}
\label{tab5} 
\end{table}

If we use the broad H$\alpha$ flux measured from the LRD stack to estimate the continuum flux at 5100 \AA~using Eq. 1 in \citet{Greene2005}, as we did for our local Sy1 galaxies in Section \ref{normalizations} (see Tables \ref{tab2} and \ref{tab3}), we derive a value of $\rm \lambda F_{5100}$=$\rm 7.95\times10^{-17}~erg~s^{-1}~cm^{-2}$. This flux is only a factor of 1.4 higher than the flux at 5100 \AA~that we measured from the LRD stack (see Table \ref{tab5}). Indeed, if we use Eq. 5 in \citet{Stern2012}, more commonly used for high-redshift and more luminous AGN, we get $\rm \lambda F_{5100}$=$\rm 5.15\times10^{-17}~erg~s^{-1}~cm^{-2}$, which is 0.91 times the flux measured directly from the continuum. This would indicate that broad H$\alpha$ is a good proxy of the optical continuum luminosity in high-redshift LRDs and therefore of the AGN bolometric luminosity, contrary to what it has been recently claimed \citep{Greene2026,Brazzini26}. If instead we use the total H$\alpha$ flux (broad+intermediate components), we get a $\rm \lambda F_{5100}$ a factor of 2.6 and 1.8 higher than that measured directly from the continuum, using the \citet{Greene2005} and \citet{Stern2012} relations, respectively. If this total H$\alpha$ flux is truly representative of the BLR emission, it might be indicating that the BLR covering factor is indeed somehow higher than that of local AGN.  

Using the 5100 \AA~luminosity reported in Table \ref{tab5}, which is measured directly from the LRD stack, and $\rm L_{bol}=9\times \lambda L_{5100}$ \citep{Kaspi2000,Shen2011}, we get $\rm log~L_{bol}=44.5$ erg~s$^{-1}$. This value is in agreement with the bolometric luminosity derived from SED fitting of the LRD stack ($\rm log~L_{bol}$=44.1-44.4 erg~s$^{-1}$; \citealt{Perez26}). If we use the broad (total) H$\alpha$ luminosity reported in Table \ref{tab5} and $\rm L_{\rm bol}=(130^{\times 2.4}_{\div 2.4})\times L_{\rm H\alpha}^b$ from \citet{Stern2012}, we measure $\rm log~L_{bol}$=44.4 (44.7) erg~s$^{-1}$. If instead we use the bolometric corrections of $\rm L_{\rm bol}=5.4\times \lambda L_{\rm 5100}$ and $\rm L_{\rm bol}=19\times L_{\rm H\alpha}$ derived for two luminous LRDs by \citet{Greene2026}, we get $\rm log~L_{bol}=44.2$ erg~s$^{-1}$ from the 5100 \AA~luminosity of the LRD stack and 43.6 (43.9) erg~s$^{-1}$ from the broad (total) H$\alpha$ luminosity reported in Table \ref{tab5}. Therefore, we find that the H$\alpha$ bolometric corrections derived from local samples of type-1 AGN reproduce the bolometric luminosity derived from SED fitting of the LRD stack by \citet{Perez26}. More specifically, they require bolometric corrections of 60-130 (30-60) if we use the broad (total) H$\alpha$ luminosity, consistent with \citet{Stern2012}.

%Using the bolometric luminosities reported in Table \ref{tab1} for our Sy1 galaxies, which come from X-ray measurements, and the broad H$\alpha$ luminosities reported in Table \ref{tab2}, we find that we need bolometric corrections in the range 110-370 for all the Sy1 galaxies but NGC\,4395, as in \citet{Stern2012}. We note, however, that the bolometric luminosities reported in Table \ref{tab1} come from X-ray measurements, and not from SED fitting as in e.g. \citet{Richards2006}. 

In the case of LORD (z=0.1), \citet{Ji2026} fitted the H$\alpha$ profile detected in their Gemini/GMOS spectra (R$\simeq$3400) with two broad Gaussians of FWHM= 730 (NB) and 2050 (BB) km~s$^{-1}$ plus a narrow Gaussian of 90 km~s$^{-1}$ (see Figure 2 and Table 1 in \citealt{Ji2026}). Adding the flux of the NB and BB components, they obtained a broad H$\alpha$ luminosity of 41.90 erg~s$^{-1}$. Using 
\citet{Stern2012} and log $\rm M_{BH}/M_{\sun}$=6.49, they reported an Eddington ratio of 0.26$\pm$0.01, similar to that of NGC\,4395 (see Table \ref{tab1}). The EW reported by \citet{Ji2026} for NB+BB H$\alpha$ is 450 \AA, similar to the LRD stack from \citet{Perez26} but still within the range of typical EWs of local AGN of similar bolometric luminosities \citep{Stern2012}. As in the case of the high-z LRD stack, the 5100 \AA~fluxes of LORD estimated from its broad (BB) and total (NB+BB) H$\alpha$ fluxes using Eq. 5 in \citet{Stern2012} are 0.9 and 1.5 times the 5100 \AA~flux directly measured from the optical continuum. If confirmed for more examples of local and high-redshift LRDs, the optical continuum of LRDs can be estimated either from the broadest component fitted to their H$\alpha$ profiles or by the total H$\alpha$ flux (broad+intermediate components) divided by a factor of $\sim$1.5-2.

Finally, considering the 3$\sigma$ upper limit on the hard X-ray luminosity of LORD, obtained by \citet{Ji2026} from NuSTAR and Chandra data (log $\rm L_{2-10 keV}<$41.9 erg~s$^{-1}$), $\rm L_{bol}/L_{2-10 keV}$>137 if we estimate $\rm L_{\rm bol}$ as 130$\rm \times L_{\rm H\alpha}^b$. If we calculate the bolometric luminosity from the flux of the BB H$\alpha$ component only, reported in Table 1 in \citet{Ji2026}, we get $\rm L_{bol}/L_{2-10 keV}$>80, and with the 5100 \AA~luminosity measured from the Gemini spectrum, of 42.87 erg~s$^{-1}$, $\rm L_{bol}/L_{2-10 keV}$>88 using $\rm L_{bol}=9\times \lambda L_{5100}$ \citep{Kaspi2000,Shen2011}. This factor is closer to the limit derived from the BB H$\alpha$ component, but they are still much larger than the local $\rm L_{bol}/L_{2-10 keV}$ values of $\sim$8-25 \citep{Ho2009,Duras2020}, due to the X-ray faintness of LRDs. In the case of the Sy1 galaxies considered here, if we use the broad H$\alpha$ luminosities reported in Table \ref{tab2} and the X-ray luminosities reported in Table \ref{tab1}, we find $\rm L_{bol}/L_{2-10 keV}$<24, with a median value of 10.5.

\section{Extinction corrections}\label{appendixD}

Here we show the same plots as in Fig. \ref{fig6}, but correcting the broad H$\alpha$ and 5100 \AA~fluxes measured from the LRD stack of \citet{Perez26} by different $\rm A_V$ values using the SMC bar extinction curve of \citet{Gordon2003}.

\begin{figure*}[!h]
  \centering
 {\par\includegraphics[width=1.0\columnwidth]{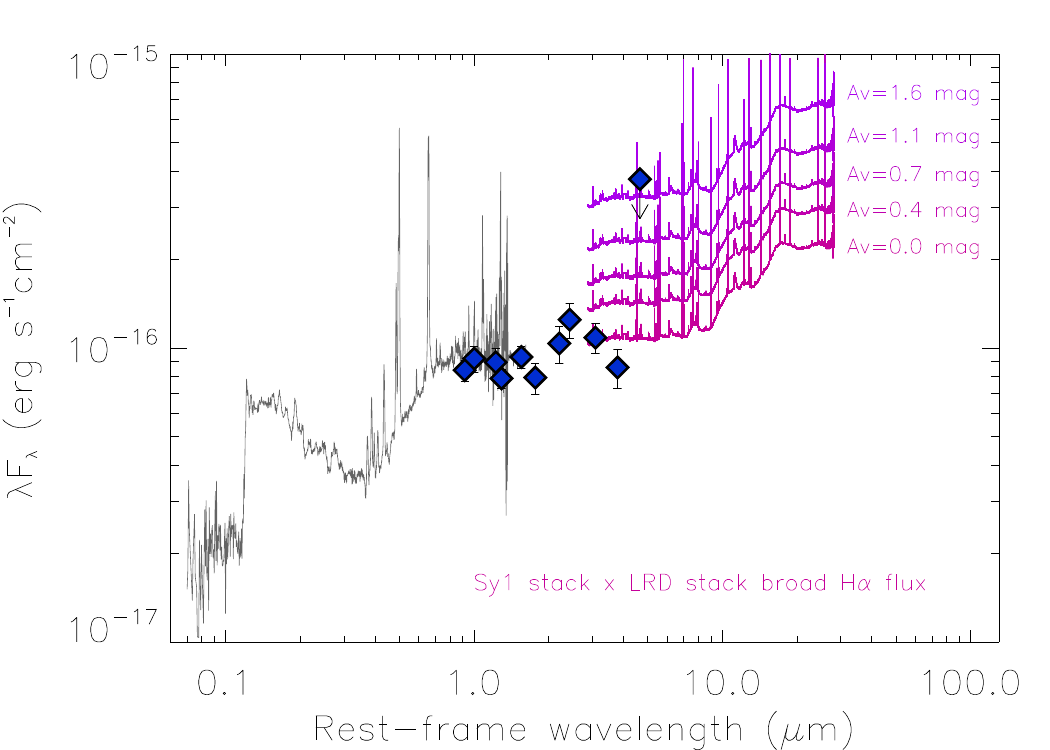}
  \includegraphics[width=1.0\columnwidth]{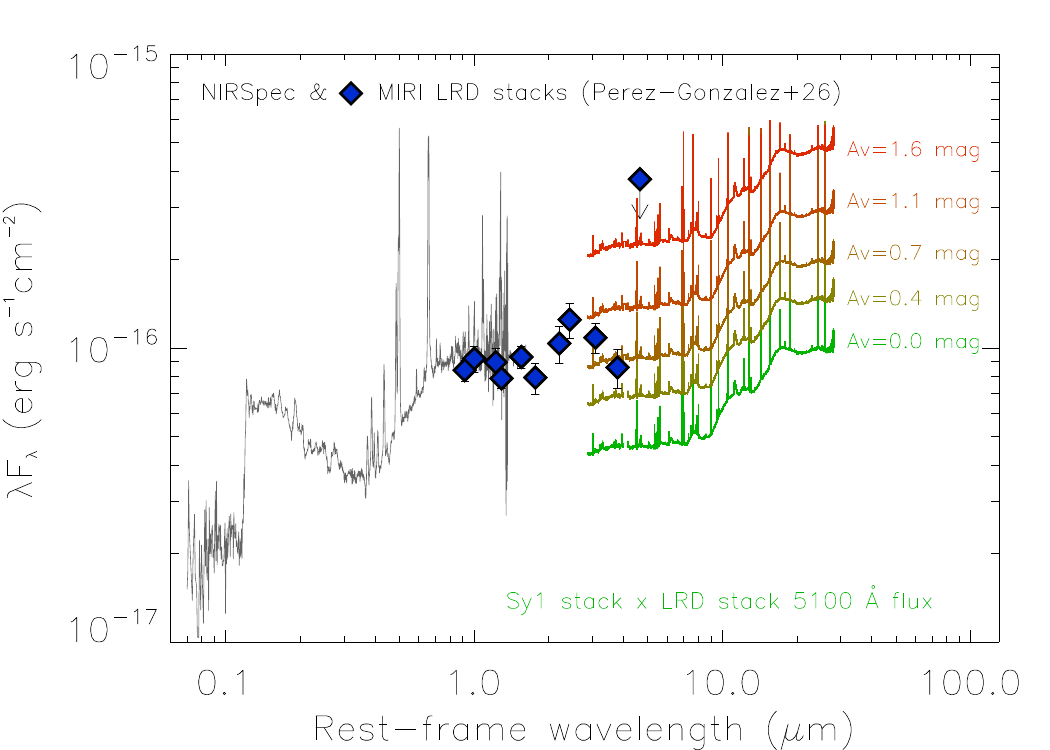}\par}
  \caption{Same as in Fig. \ref{fig6}, but correcting the H$\alpha$ and 5100 \AA~fluxes measured from the LRD stack of \citet{Perez26} by different $\rm A_V$ values using the SMC bar extinction curve of  \citet{Gordon2003}. The Sy1 stacks have also been corrected from the same values of the extinction, but using the local ISM extinction law of \citet{Chiar06}.}
              \label{figD1}
    \end{figure*}

\section{Dust masses}\label{appendixE}

\begin{figure}[!h]
  \centering
\includegraphics[width=1.0\columnwidth]{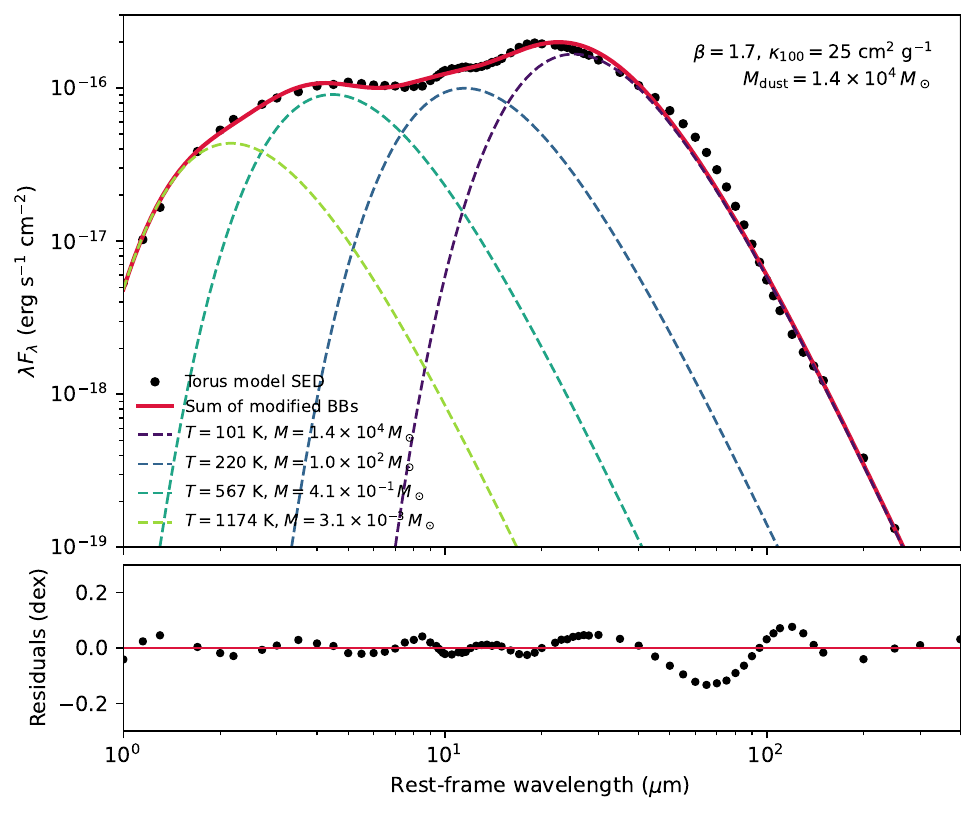}
  \caption{Torus model shown in Fig. \ref{fig4} (black dots) and best-fit sum of four optically thin modified blackbodies ($\beta=1.7$, $\kappa_{100}=25$\,cm$^2$\,g$^{-1}$; red solid line). The dashed lines show the individual components. The bottom panel shows the residuals in dex.}
              \label{fig:BBs}
    \end{figure}

We fitted the dust torus model in Fig. \ref{fig4} with a sum of four modified BBs ($L_\nu=4\pi M_{\rm dust}\kappa_\nu B_\nu(T)$, with $\kappa_\nu=\kappa_{100}\,(100\,\mu{\rm m}/\lambda)^{\beta}$; see Fig.~\ref{fig:BBs}). We adopted $\beta=1.7$, as constrained from the slope of the long-wavelength tail, from $\sim$200 to 5000 $\mu$m ($\lambda F_\lambda\propto\lambda^{-4.65}$), and $\kappa_{100}=25$\,cm$^2$\,g$^{-1}$ \citep{Draine2003}. The fit reproduces the torus SED at $\lambda\geq1\,\mu$m with an rms of 0.05\,dex. Using three BBs instead of four produces an rms of 0.07\,dex. More than 99\% of the dust mass is in the coolest component ($T\simeq100$\,K), giving $M_{\rm dust}\simeq1.4\times10^4\,M_\odot$ ($1.3\times10^4\,M_\odot$ using three BBs instead of four).

\end{appendix}

%\end{appendix}

\end{document}